\documentclass[aps,prx,pdflatex,twocolumn,superscriptaddress,longbibliography]{revtex4-1}
\usepackage[compat=1.0.0]{tikz-feynman}
\pdfoutput=1

\usepackage{amssymb}
\usepackage{multirow}
\usepackage{bm}
\usepackage{amsmath}
\usepackage{graphicx}
\usepackage{epstopdf}
\usepackage{subfigure}
\usepackage{natbib}
\usepackage{epsfig}
\usepackage{amsfonts}
\usepackage{mathrsfs}
\usepackage[toc,page,title,titletoc,header]{appendix}
\usepackage[colorlinks,linkcolor=blue,citecolor=blue,anchorcolor=blue]{hyperref}

\usepackage{braket}

\usepackage{dsfont,amsthm,amsbsy}

\usepackage{fancyhdr}

\usepackage{ulem}

\usepackage{bbm}

\newcommand{\bea}{\begin{equation} \begin{aligned}}
\newcommand{\eea}{\end{aligned} \end{equation} }

\newcommand{\bpm}{\begin{pmatrix}}
\newcommand{\epm}{\end{pmatrix}}

\def\spinc{spin$_\mathbb{C}$ }

\begin{document}

\title{Anyon superfluidity of excitons in quantum Hall bilayers}

\author{Zhaoyu~Han}
\affiliation{Department of Physics, Harvard University, Cambridge, Massachusetts 02138, USA}

\author{Taige~Wang}
\affiliation{Department of Physics, University of California, Berkeley, CA 94720, USA}

\author{Zhihuan~Dong}
\affiliation{Department of Physics, University of California, Berkeley, CA 94720, USA}

\author{Michael~P.~Zaletel}
\affiliation{Department of Physics, University of California, Berkeley, CA 94720, USA}

\author{Ashvin~Vishwanath}
\affiliation{Department of Physics, Harvard University, Cambridge, Massachusetts 02138, USA}

\begin{abstract}

The charged anyons of a fractional quantum Hall fluid are necessarily dispersionless due to the continuous magnetic translation symmetry. Neutral anyons, however, can disperse, resulting in a much richer space of possible ``daughter'' states when doped to finite density. We discuss a natural realization of such physics in quantum Hall bilayers, where a finite density of excitons with fractional statistics is argued to give rise to `anyonic exciton superfluidity,' the charge-neutral analog of anyon superconductivity. In a balanced bilayer of two Laughlin $\nu = 1/3$ states, the minimal interlayer exciton carries anyonic exchange statistics. A finite density of these excitons is argued to yield an exciton superfluid stitched to a specific bulk topological order and edge spectrum. Such superfluidity should be most robust near the direct transition into the Halperin $(112)$ state, and near analogous transitions in the bilayer Jain sequence  at total filling $\nu_\text{T} = 2\times \frac{n}{2n+1}$. These topological transitions can be  described by Chern-Simons QED$_3$, from which we derive several novel and general properties of anyon superfluidity near such transitions, including an anomalously large superfluid stiffness of $\kappa_\text{s} \propto |\delta\nu|^{1/2}$ at layer imbalance fraction $\delta\nu$. A notable feature of the phase diagrams we construct is the prevalence  of spatial symmetry breaking, driven by an underlying composite Fermi surface. Our results can be directly tested with currently available experimental techniques. We compare our theory with existing data and make concrete predictions for future measurements, including higher-pseudospin exciton superfluids when doping higher Jain fractions.
\end{abstract}

\maketitle

\section{Introduction}

The study of quantum Hall (QH) effects in two-dimensional electron systems subjected to strong magnetic fields has unveiled a wealth of novel quantum phases with fractionalized anyonic excitations. Despite the richness of the phases, the dynamics of excitations therein is typically limited owing to the strong fields. There are two distinct scenarios where such quenching is avoided. First, when the continuous magnetic translation symmetry is disrupted by a periodic potential, the quasi-particles develop a dispersion, resulting in intriguing outcomes, such as quantum phase transitions between different states~\cite{PhysRevB.56.R7100,PhysRevX.8.031015,PhysRevLett.125.236805,PhysRevResearch.2.033348,PhysRevB.109.085143} and conducting daughter states upon charge doping~\cite{shi2024doping,divic2024anyon,nosov2025plateau,shi2025anyon,pichler2025,wang2025}. This line of inquiry has gained momentum recently with the discovery of the fractional quantum anomalous Hall effect~\cite{cai2023signatures,PhysRevX.13.031037,lu2024fractional,xie2021fractional} and, more broadly, QH states in moiré superlattices~\cite{doi:10.1126/science.aan8458,doi:10.1126/science.adi4728,kometter2023hofstadter,yu2022correlated,saito2021hofstadter,doi:10.1126/science.aad2102}. Second, even with perfect translation symmetry, the charge-{\it neutral} quasi-particles can traverse freely, unimpeded by the magnetic field. In this work, we will investigate this scenario by focusing on a facile platform - QH bilayers - where we show the neutral anyons, endowed with a layer pseudospin, can exhibit non-trivial dynamics and lead to collective effects including quantum phase transitions between distinct QH states and the excitonic analog of anyon superconductivity~\cite{doi:10.1126/science.242.4878.525,PhysRevLett.60.2677,PhysRevB.42.342,WilczekWittenHalperinAnyonSC,PhysRevB.39.11413,PhysRevLett.62.2873,PhysRevB.39.9679,PhysRevLett.63.903,PhysRevB.41.240,PhysRevB.103.165138,shi2024doping,divic2024anyon}.

QH bilayers host a rich family of topological phases tuned by layer seperation and perpendicular fields~\cite{annurev:/content/journals/10.1146/annurev-conmatphys-031113-133832,YE2008580,liu2019interlayer,doi:10.1126/science.aao2521,PhysRevLett.86.1849,PhysRevB.47.16419,PhysRevLett.64.1313,PhysRevB.86.035326,PhysRevB.93.085436,PhysRevB.110.195106,PhysRevLett.127.246803,PhysRevX.13.031023,PhysRevLett.132.176502,PhysRevResearch.5.L042022,PhysRevLett.91.046803,PhysRevLett.87.056802,PhysRevLett.101.176803,PhysRevLett.119.177601,PhysRevLett.88.106801,PhysRevB.78.195327,PhysRevB.82.235312,PhysRevLett.120.077601,PhysRevB.95.085135,PhysRevB.82.233301,PhysRevB.65.041305,PhysRevLett.101.176803,PhysRevLett.105.216804,PhysRevB.81.045323,PhysRevB.98.045113,PhysRevLett.121.026603,li2019pairing,zhang2025excitons,nguyen2025bilayerexcitonslaughlinfractional,zeng2023evidencesuperfluidtosolidtransitionbilayer,PhysRevLett.117.096803,PhysRevB.84.115121,PhysRevB.92.035103,PhysRevLett.113.236804}. In their simplest setup, the electron spins are fully polarized and the two layers are well insulated from each other, so that the layer index, or pseudospin, $\sigma \in \{\uparrow, \, \downarrow\} = \updownarrow$, labels the only two separately conserved components of the electrons, $\psi^{(\sigma)}$. Tuning the ratio between the distance between the two layers, $d$, and the magnetic length of the out-of-plane magnetic fields, $\ell_B$, can drive transitions between distinct states. Intuitively, when $d/\ell_B $ is large, the two layers are decoupled, and should form two separate single-component QH states corresponding to their filling fractions, $\nu^{(\sigma)}$; while as $d/\ell_B$ is lowered, the electrons experience stronger inter-layer Coulomb repulsion, and more complicated states can be expected. Another important control parameter is the layer–density imbalance, i.e., the pseudospin polarization, $\delta\nu = (\nu^{(\uparrow)} -\nu^{(\downarrow)})/2$. Tilting $\delta \nu$ while holding the total filling $\nu_\text{T}= \nu^{(\uparrow)} +\nu^{(\downarrow)}$ fixed injects charge-neutral excitons into the QH fluid. 

A central question is: What ground state emerges when a finite density of pseudospin is introduced? The outcome hinges on the characters of the lowest–energy neutral excitation that carries pseudospin. If it is an ordinary electron–hole bound state residing in opposite layers, its Bose condensation yields an exciton superfluid (ES) coexisting with the original topological order. Far richer possibilities arise, however, if the pseudospin carrier is fractionalized—i.e. an anyon endowed with non-trivial self-statistics. Doping such anyonic excitons can again produce an ES, by the same mechanism that underlies anyon superconductivity~\cite{PhysRevLett.60.2677}.

As a concrete example, let's consider the balanced filling $\nu^{(\updownarrow)} = 1/3$, near which various signatures of exciton formation, such as Coulomb drag, have been experimentally identified in the pseudospin-doped regime~\cite{zhang2025excitons,zeng2023evidencesuperfluidtosolidtransitionbilayer,nguyen2025bilayerexcitonslaughlinfractional}. At this filling, there are two natural candidate states constructed by Halperin~\cite{Halperin:1983zz}, the so-called $(330)$ and $(112)$ states which have distinct Abelian topological orders as well as different pseudospin and geometric responses~\cite{Wen01101995}. The $(112)$ state has an inter-layer phase winding structure that implies a larger average inter-particle distance between electrons on the opposite layer than the $(330)$ state which is essentially two decoupled layers of Laughlin $\frac{1}{3}$ states. So as $d/\ell_B$ decreases, a transition from the latter to the former state is anticipated. This has indeed been observed in a density-matrix-renormalization-group study~\cite{PhysRevB.91.205139} and an earlier exact diagonalization study~\cite{PhysRevB.53.15845}. The former study interpreted the transition to be weakly first order; in an upcoming work~\cite{numerics}, we will present numerical evidence showing that the transition can be continuous. The observations in a recent experiment~\cite{nguyen2025bilayerexcitonslaughlinfractional} seem to be consistent with this conclusion, which show a feature in conductance measurements at a value of $d/\ell_B$, indicating a critical point. In this experiment, similar transition-like features are also identified at filling fractions $\nu^{(\updownarrow)} = 2/5$ and $ 3/7$, which, along with $\nu^{(\updownarrow)} = 1/3$, belong to the principal Jain sequence~\cite{PhysRevLett.63.199} $\nu^{(\updownarrow)} = \frac{n}{2n+1}$ ($n\in\mathbb{Z}$). These observations lead to the following theoretical questions: 
\begin{itemize}
    \item What is the effective description of the continuous transitions -- if they exist at all? 
    \item What implications do the transitions have on the properties of the ES obtained upon doping pseudospin?
\end{itemize}

In this work, we address these two questions within a two-band model of a charge-neutral and pseudospin-carrying anyon, which is derived from a two-component composite fermion (CF) theory~\cite{Jain_2007}. We will show that the Chern transition of the bands will naturally give rise to a QED$_3$-Chern-Simons description to the criticality, where these anyons acquire Dirac dispersion. Based on this theory, we will also outline a mechanism for ES upon doping these anyons, which parallels that proposed for anyon superconductivity and thus we refer to as anyonic ES. We argue that such anyon superfluidity is  energetically favorable near the transition point, because the relevant anyons become simultaneously `cheap' and `light' there, i.e. possessing vanishing Dirac and Newton (inertial) mass. Indeed, we find that the anyonic ES state we construct has an anomalously soft energy scaling $\sim|\delta\nu|^{3/2}$ and large superfluid stiffness $\sim|\delta\nu|^{1/2}$ at small layer-imbalance $\delta\nu$, demonstrating its stability against thermal fluctuations and also potentially against possible competing phases and disorder effects. The anyon ES mechanism also makes specific predictions for the properties of the ES, including the pseudospin and angular momentum of the elementary exciton in the condensate, the co-existing topological order, and the edge states. { We will present numerical evidences supporting our results in an upcoming work~\cite{numerics}.} In particular, we note that the theory predicts pseudospin-$2n$ ES for filling fraction $\nu^{(\updownarrow)} = \frac{n}{2n+1}$, potentially opening new route to the realization of ``spin nematic'' order emerging from higher Jain fractions. Our results suggest that QH bilayers can serve as novel and controllable arena for the study of anyon superfluidity.

\begin{figure}[t]
    \centering
    \includegraphics[width=0.8\linewidth]{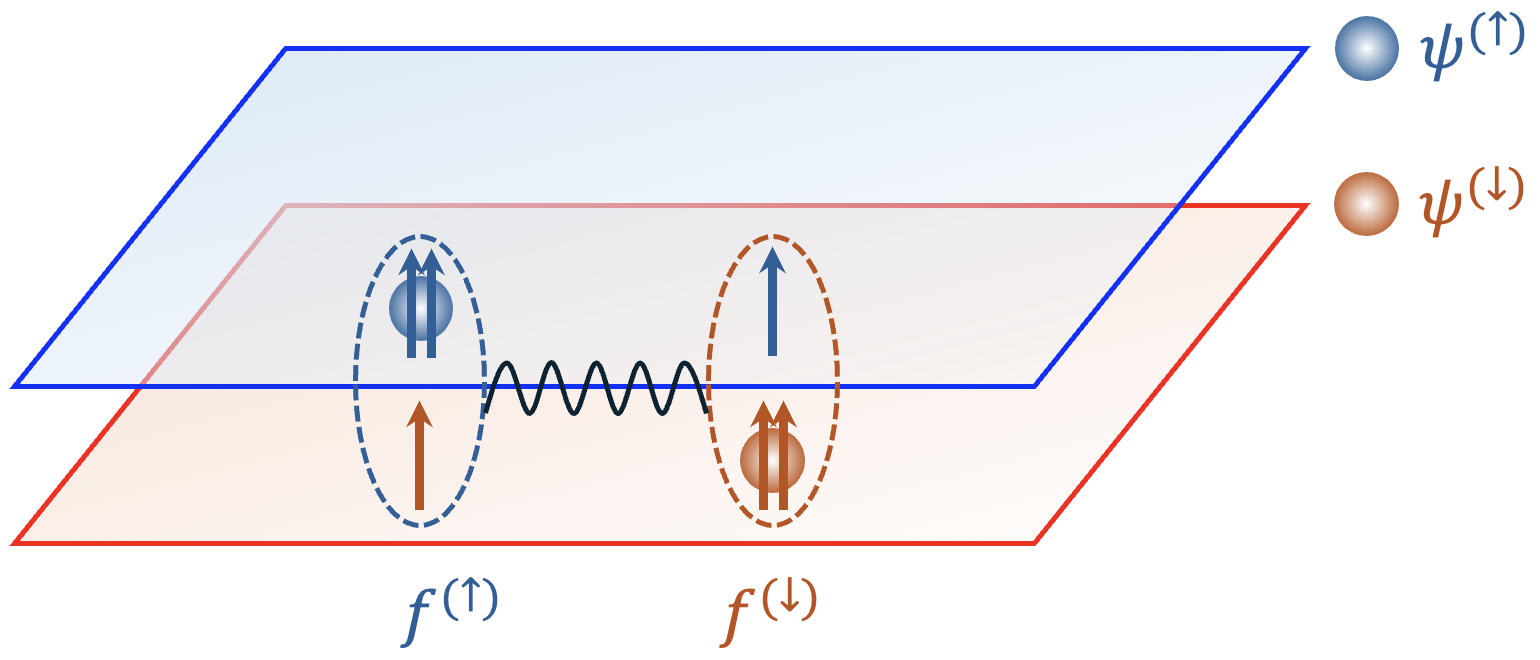}
    \includegraphics[width=\linewidth]{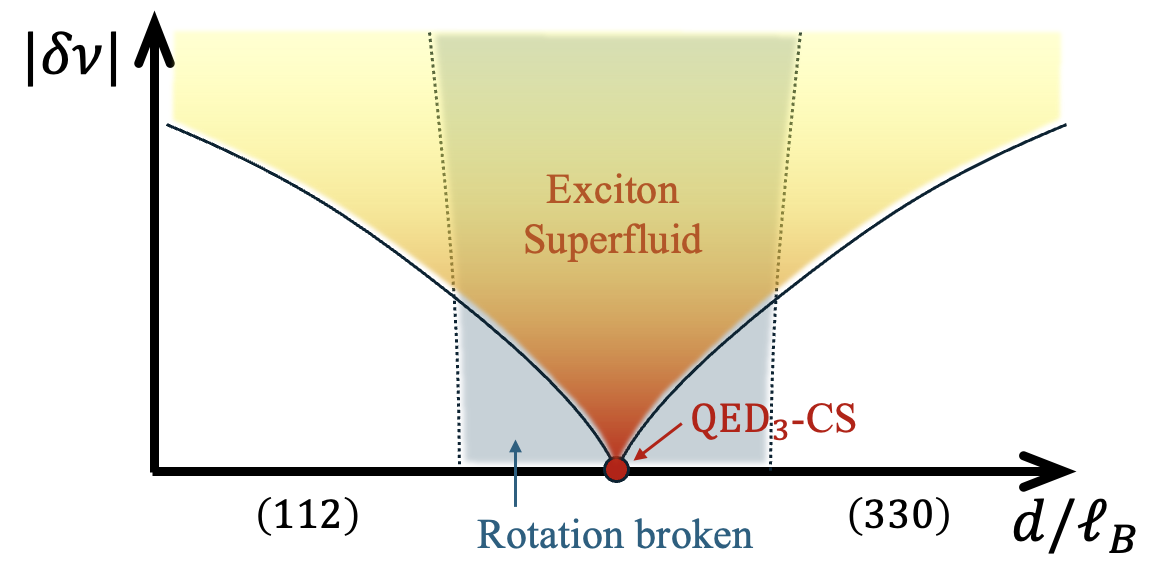}
    \caption{(Top) An illustration of the flux attachment scheme of the composite fermion theory for $\nu_\text{T} =2\times 1/3$, and the inter-layer pairing scenario (wavy line). Each arrow represents a $2\pi$ flux within the layer.  (Bottom) A proposed schematic phase diagram near the transition out of the layer-decoupled $(330)$ state by tuning $d/\ell_B$, the layer seperation,  and $\delta\nu$, the pseudospin filling fraction defined by $\nu^{(\sigma)} =1/3 +\sigma \cdot \delta\nu$. The uncolored regions have multiple possibilities, including QH plateau states in which the doped quasi-particles are pinned by the formation of a crystal or by disorder. In the shaded region, the $SO(2)$ rotation symmetry must be broken or additional topological order must be introduced, in order to allow for the continuous transition point. In the absence of rotation symmetry larger than $C_2$, the ESs at different $d/\ell_B$ could belong to the same phase. For the  $\nu_\text{T} =2\times 1/3$ case the ES has pseudospin-$2$, while for higher Jain fractions $\nu^{(\updownarrow)} = \frac{n}{2n+1}$, our theory suggests a pseudospin-$2n$ ES emanating from the critical point.}
    
    \label{fig: schematic}
\end{figure}

\section{Overview}

\label{sec: overview}

Here we first outline our essential results, highlighting the physical picture and directing the reader to specific sections in the main body of the paper where more details can be found. We will primarily discuss the simplest case with $\nu^{(\updownarrow)}=1/3$ and the main results will be generalized to other Jain fractions.

We start with a general discussion on charge-neutral excitations that carry pseudospin in QH bilayers. The simplest such excitation is an exciton -- an electron-hole pair. This is a trivial, electrically neutral boson whose condensation leads to ES and leaves the original topological order, if any, intact. There could also be anyonic excitons formed by pairs of anyons with opposite electric charge on the two layers, which can possess fractionalized statistics and pseudospin. When these are the lowest-energy pseudospin carrying (but charge neutral) excitations, they will end up being doped into the system on tuning layer imbalance. Certain types of such anyonic excitons are still self-bosons, but  braid non-trivially with certain other anyons. They can therefore directly condense, producing an ES phase which confines those other anyons in the topological order. Such a scenario has been predicted, for instance, in the  layer-imbalanced fraction $\nu^{(\updownarrow)}=1/3,2/3$, with $\nu_\text{T} =1$ setup~\cite{PhysRevX.13.031023} where signatures of excitons have been reported experimentally upon lowering $d/\ell_B$~\cite{zeng2023,nguyen2025bilayerexcitonslaughlinfractional}. Excitons could also exhibit fermionic self-statistics,  doping these fermionic excitons is expected to produce an  exciton Fermi liquid, a phase proposed in certain QH bilayer systems~\cite{PhysRevLett.121.026603,PhysRevB.98.045113}. 

Finally, we come to the most interesting case where the doped excitons carry {\it non-trivial} self-statistics, i.e. are anyonic. This turns out to be the scenario relevant to the current discussion, and the mechanism is more involved than in the previous instances. As a concrete example - central to this paper - we focus on a balanced bilayer with fillings $\nu_\uparrow=\nu_\downarrow= 1/3$.  The simplest interlayer bound state is the $\vec{Q} = (1/3, -1/3)e$ composite: a Laughlin quasiparticle in one layer bound to a quasihole in the other layer.  This `fractional exciton' is electrically neutral but carries pseudospin $S = 2/3$ (taking an electron–hole pair to have $S = 2$).  Because each constituent Laughlin quasiparticle has statistical angle $\pi/3$, and the layers are independent, the composite inherits anyonic self-statistics $\theta = 2 \times (\pi/3)$. Introducing a small pseudospin imbalance then creates a finite-density gas of these anyonic excitons.  What state of matter does this gas form?  The question parallels the classic `anyon superconductivity' problem~\cite{doi:10.1126/science.242.4878.525,PhysRevLett.60.2677,PhysRevB.39.9679,WilczekWittenHalperinAnyonSC}.  Extending those arguments, we find that the excitons condense into an excitonic superfluid, that coexists with a specific bulk topological order and corresponding edge modes.

The intuition is the following. View the anyonic exciton with self statistics $\theta = 2\pi/3 =\pi -\pi/3$,  as a fermion carrying an attached statistical flux $-\pi/3$ that reproduces the anyonic exchange statistics.  In a mean-field picture, smearing this flux into a uniform background produces an effective magnetic field under which the fermions are at filling $\nu_f = -3$, i.e., three filled Landau levels.  Fluctuations that promote the statistical gauge field to a dynamical one then transmute this integer-quantum-Hall state of composite fermions into an exciton superfluid.  The construction is readily seen to be internally consistent: the phase remains compressible for instance, because any change in exciton density simultaneously adjusts both the fermion and flux densities to keep the effective filling fixed. See App.~\ref{app: smith} for a mathematical   formulation of this picture.

The only physical difference from the Cooper pair analog is that here the repulsive Coulomb interaction directly favors the formation of anyonic excitons with the requisite self-statistics for anyon superfluidity; in this perspective, the QH bilayer setting provides a more stable and controlled arena where we can study aspects of the anyon superconductivity mechanism.  Furthermore, in our treatment the excitonic superfluid is predicted to coexist with a specific topological order and edge state content.

Let us now further consider $\nu^{(\updownarrow)}=1/3$, as the separation between layers is decreased. This problem is already interesting even in the absence of pseudospin doping, on account of the  $(330)$-$(112)$ transition. Importantly, the $(330)$ and $(112)$ states have identical charge (parallel-flow) Hall response but distinct pseudospin (counterflow) Hall response, directly accessible in counterflow transport. Denoting the charge and pseudospin Hall conductance by $\sigma_{xy}^{\mathrm{c}}$ and $\sigma_{xy}^{\mathrm{s}}$, respectively, we have (in units $h/e^{2}=2\pi$)
\begin{equation}
    2\pi \sigma_{xy}^{\mathrm{c}}=\frac{2}{3},\qquad 
2\pi \sigma_{xy}^{\mathrm{s}}=
\begin{cases}
2/3, & (330),\\[4pt]
-2, & (112).
\end{cases}
\end{equation}
Throughout the $(330)$–$(112)$ transition the charge gap remains open while the pseudospin gap must close to allow for the change in counterflow Hall conductance. Experimentally, this implies that the longitudinal conductance in the parallel-flow channel remains negligible $\sigma_{xx}^{\mathrm{c}}\!\approx 0$, whereas the counter-flow longitudinal conductance $\sigma_{xx}^{\mathrm{s}}$ increases on approaching criticality. Furthermore, the change in topological order is also limited to the pseudospin sector. We will adopt a composite fermion (CF) description of the relevant anyonic excitons to obtain a global phase diagram (sketched in Fig.~\ref{fig: schematic}). Within this theory, the $(330)$-$(112)$ transition can be readily explained by a gap closing transition of the anyon band which reduces the topological order in the pseudospin sector. Near the transition, these anyonic excitons thus soften relative to all other excitations.  Again in this broader setting with an effective CF description, these low-energy anyons can lead to an exciton superfluid when pseudospin is doped.

These resulting ES phases upon doping both $(330)$ and $(112)$ states have the same residual topological order, which is essentially equivalent to the $(112)$ state. The fact that the $(112)$ topological order does not change on doping is to be expected, since in this state there is no non-trivial anyonic exciton excitation. Hence, after doping, both large- and small-$d/\ell_B$ limits share the same anyon content plus an ES order.  Are they therefore the same phase?  The answer hinges on symmetry.  More specifically the `shift,' a symmetry enrichment of topology, represents a universal property of the phase that depends on charge conservation symmetry and spatial rotation symmetry~\cite{PhysRevLett.69.953}. Because the parent $(330)$ and $(112)$ states possess different shift, and because the shift remains unchanged during pseudospin doping on account of preserving the charge gap, it is easily determined that the ES that descend from the two zero-doping phases also differ in their shift. If full rotational symmetry is preserved, a quantum phase transition must separate the two excitonic superfluids.  Alternatively, if rotation symmetry is explicitly or spontaneously broken to $C_2$ (or lower), the transition can be avoided and the two limits can fuse into a single phase.

In the latter direct transition scenario, we can address the second question - is there a mechanism for enhancing superfluidity? In this exciton setting, pairing is already granted, the bottleneck is the pseudospin phase stiffness $\kappa_\text{s}$, which necessarily vanishes as the pseudospin density vanishes $\delta \nu \rightarrow 0$. Therefore, the key question then is the functional form of the dependence of stiffness on density. We argue that while doping the phases typically leads to $\kappa_\text{s} \propto |\delta \nu|$, we expect a parametrically {\it enhanced} stiffness in the vicinity of a critical point with conformal symmetry: $\kappa_\text{s} \sim |\delta\nu|^{1/2}$. Criticality thus acts as a catalyst producing a stiffer and hence more robust superfluid. Note that this is distinct from recent work~\cite{divic2024anyon} where topological criticality was key also to inducing pairing - however we note that in addition there, conformal criticality should also lead to enhanced superfluid stiffness at the transition.

Below we provide navigation to the key steps:

{\bf Composite Fermions:} To capture the various phases on an equal footing, we introduce CFs $f^{(\sigma)}$, obtained by attaching two flux quanta in the same layer and one in the opposite layer to each electron (Sec.~\ref{sec: CF theory}).   At filling $\nu^{(\updownarrow)}=1/3$ these CFs see zero effective field and form two Fermi surfaces, one per pseudospin. The incompressible QH states of interest then correspond to inter-layer paired superconductors of CFs (Sec.~\ref{sec: pairing}): the simple weak pairing $p_x\pm \mathrm{i} p_y$ phases of CFs remarkably correspond to the $(330)$ and $(112)$ Halperin states, respectively (Sec.~\ref{sec: p pm p}). 

{\bf Transition at zero doping:} To transition between these states, consider weakening one of the components (say the $p_y$ component) and changing its sign, which creates a pair of Dirac points on the Fermi surface (Sec.~\ref{sec: transition p pm p}). Their arbitrary positions signal broken rotational symmetry, already satisfied if rotation symmetry is broken prior to reaching the critical point, either by hand or spontaneously (Sec.~\ref{sec: symmetry}). Note, all we need is for an axis to be selected, which implies that it is sufficient to break rotation down to $C_2$.  At criticality, the presence of two Dirac points separated { by a momentum $\approx 2k_F$, where $2k_F$  is} the size of the composite fermion Fermi sea, implies enhanced pseudospin density wave correlations with this period. 

{\bf Doping pseudospin:} This is most conveniently analyzed by considering the paired CFs in the Nambu basis $F=(f^{(\uparrow)},-f^{(\downarrow),\dagger})^T$. This spinor field describes exactly the relevant anyonic excitons, whose band Chern number jumps across the above mentioned transition. Doping pseudospin injects $F$ and the flux they see in proportion $2:1$, reminiscent of the case in anyon superconductivity. When this leads to filled Landau levels (LL) for $F$, the resulting electronic state is the discussed anyon ES (Sec.~\ref{sec: kinematics}). An alternate derivation of the topological properties of the resulting ES state can be directly obtained using a recently developed stack-and-condense approach for anyons~\cite{zhang2025hierarchyconstructionnonabelianfractional,shi2025}, which we apply to our case of interest in App.~\ref{app:sec: stack&condenseHierarchy}.

Near criticality, the Dirac theory of $F$ makes a further prediction -- in order to achieve the proper  LL filling factor, a valley ordering is necessary. This forces additional breaking of rotational or translational symmetry, a sharp prediction for future numerical or experimental studies. Also, we find criticality can enhance phase stiffness, specifically through its functional dependence on doping. These features can serve as `dynamical' hallmark of anyon superfluidity in addition to `kinematics' signatures including the residual topological order, edge states and shift quantum number of the exciton condensate (Sec.~\ref{sec: mechanism}). 

{\bf Generalizations:} Our theory is readily generalized to other fractions in the bilayer Jain sequence $\nu^{(\updownarrow)} = n/(2n+1)$. The CF theory is generalized to a fermionic parton theory in those cases, and again the two possible undoped phases are obtained by consider their paired states. We compare the expectations within this framework to earlier experimental observations~\cite{liu2019interlayer} at a relatively small $d/\ell_B$. For such a setting, we predict that doping in the vicinity of the transition will lead to pseudospin-$2n$ ES (Sec.~\ref{sec: generalization}).

\section{Formalism}
\label{sec: formalism}

The systems of interest have two separate charge $U(1)_{\uparrow}\times U(1)_{\downarrow}$ conservation on the two layers, or equivalently charge and pseudospin {  $[U(1)_c\times U(1)_s]/\mathbb{Z}_2$ conservation, which allow the coupling to two external gauge fields,  $A^{(\sigma)}$ for the two layers ($\updownarrow\equiv\pm$ in equations), or equivalently charge and pseudospin gauge fields $A^{\text{c}}\equiv  (A^{(\uparrow)} + A^{(\downarrow)})/2$ and $A^{\text{s}}\equiv  (A^{(\uparrow)} - A^{(\downarrow)})/2$ (note that they are they are not independently quantized $U(1)$ gauge fields due to an additional $\mathbb{Z}_2$ quotient identifying $\pi$-phase rotations of spin and charge, but this property will not be important for transport properties we are analyzing below).} The background magnetic field $\nabla\times \bm{A}^{(\sigma)} =\bar{B}$ is constant and the same across the two layers and the densities $\rho^{(\sigma)} \equiv |\psi^{(\sigma)}|^2 = \nu^{(\sigma)} \frac{\bar{B}}{2\pi}$ are independently tunable by gates.

To simplify the below discussions and notations, we will further assume an interlayer exchange symmetry $\mathcal{X}:\uparrow \leftrightarrow \downarrow$ unless an explicit bias voltage between the two layers is applied. We will also assume $SO(2)$ spatial rotational invariance, such that the orbital spins of the particles and the geometrical responses of the QH fluids can be discussed within the Wen-Zee framework~\cite{PhysRevLett.69.953,Wen01101995}. However, we emphasize that many key results of this work do {\it not} depend on these two symmetries but only require two-fold rotation symmetry $C_2\in SO(2)$ (or equivalently the in-plane inversion $\mathcal{I}$); detailed roles of the $\mathcal{X}$ and rotation $SO(2)$ symmetry will be discussed in App.~\ref{sec: symmetry}.

Throughout this paper, we will adopt the following convention: Greek letters ($\mu,\nu,\dots$) are used for spacetime indices, and Latin letters ($i,j,\dots$) are used only for the spatial indices; the sign convention is such that $a_\mu = (a_t,\bm{a}_i)$, $\partial_\mu = (\partial_t, - \bm{\nabla}_i)$ and the metric $g_{\mu\nu} = \text{diag}(1,-1,-1)$ for flat spacetime. Bold symbols are reserved for the spatial components of physical vectors, while arrow symbols are used for abstract vectors. Repeated indices in equations indicate implicit summations over them. We absorb a factor of electron charge $|e|$ in the definition of gauge fields, and adopt units such that $\hbar = c = 1$. 

\subsection{Composite fermion theory for $\nu_\text{T} = 2/3$ }
\label{sec: CF theory}
After setting the stage, we are now ready to introduce the CF theory which applies to the total filling fraction $\nu_\text{T} = 2/3$. A fermionic parton derivation of this theory, which systematically generalizes it to other filling fractions of interest, can be found in App.~\ref{app: parton}. It can be viewed as a two-component generalization to Lopez-Fradkin~\cite{PhysRevB.44.5246} or Halperin-Lee-Read~\cite{PhysRevB.47.7312} theory of CF, described by the Lagrangian density:
\begin{align} \label{eq: L}
\mathcal{L} =&\sum_{\sigma=\updownarrow} \mathcal{L}_{f}^{(\sigma)}    + \mathcal{L}_\text{CS} + \mathcal{L}_\text{int}  \\
\mathcal{L}^{(\sigma)}_{f}/\sqrt{g}  =& f^{(\sigma),\dagger} \left[ \mathrm{i} D^{(\sigma)}_0  + \mu^{(\sigma)}  - \frac{ D^{(\sigma),i} D^{(\sigma)}_i }{2m} \right]f^{(\sigma)} \label{eq: Lf}\\
 \mathcal{L}_\text{CS}  =& \sum_\sigma \frac{1}{2\pi} \left(-a^{(\sigma)}+ A^{(\sigma)}+\omega\right) \mathrm{d}\alpha^{(\sigma)} \nonumber\\
 & \ \ \ \ \ \ - \frac{1}{4\pi}\begin{pmatrix}
\alpha^{(\uparrow)} & \alpha^{(\downarrow)}
\end{pmatrix} \begin{bmatrix}
2 & 1 \\
1 & 2
\end{bmatrix} \begin{pmatrix}
\mathrm{d}\alpha^{(\uparrow)} \\ \mathrm{d}\alpha^{(\downarrow)}
\end{pmatrix} \label{eq: LCS}
\end{align}
where we have adopted a short-handed notation for exterior derivatives, e.g. $a\mathrm{d}b = \epsilon^{\mu\nu\eta}a_\mu \partial_\nu b_\eta$ and defined covariant derivatives
\begin{align}
    D^{(\sigma)} \equiv \partial -\mathrm{i} a^{(\sigma)}.
\end{align}
$f^{(\sigma)}$ are the CF fields with two pseudospins, $a^{(\sigma)}$ and $\alpha^{(\sigma)}$ are emergent dynamical $U(1)$ gauge fields that introduce necessary flux attachment through the Chern-Simons (CS) terms in $\mathcal{L}_\text{CS}$. $\mu^{(\sigma)}$ are the chemical potentials for the electrons with two pseudospins, which are manually separated from $A^{(\sigma)}_0$ for clarity and can be independently tuned by gate voltages to ensure the desired filling conditions. $m$ is the effective mass of the CFs. To probe the gravitational and angular momentum responses, we formally introduced the coupling to the background geometry through the metric $g$ and the Abelian $SO(2)$ spin connection $\omega$~\footnote{The coupling coefficients can be determined by using the prescription in Ref.~\cite{PhysRevB.90.115139}, or by the parton approach in App.~\ref{app: parton}. We will always consider flat time direction, so $\omega = (0,\bm{\omega})$ has no zeroth component and can be defined by a set of veilbein vectors $\bm{e}^{a=1,2}$ (local coordinate basis satisfying $g_{ij} = \bm{e}^{a}_{i}\bm{e}^{a}_j$ and $\epsilon_{ij} = \epsilon_{ab}  \bm{e}^{a}_{i}\bm{e}^{b}_j$) through $\bm{\omega}\equiv \frac{1}{2}\epsilon_{ij}  \bm{e}^{b}_i \bm{\nabla} \bm{e}^{b}_j  $. $\nabla\times \bm{\omega} = R$ is the curvature, and $\frac{1}{2\pi}\int R = 2(1-g)$ on a closed manifold where $g$ is the genus. Therefore, the minimal flux of $\omega$ is $4\pi$ on closed manifolds, and the orbital spins - the `charge' under $\omega$ - are quantized in units of half integers. }. 

The flux attachment scheme can be seen from the equations of motions of $a^{(\sigma)}$ and $\alpha^{(\sigma)}$, which yield expressions for the currents for the electrons/CFs:
\begin{align} 
 \begin{pmatrix}
J^{(\uparrow),\mu} \\
J^{(\downarrow),\mu}
\end{pmatrix}  
   =& \frac{\epsilon^{\mu\nu\eta} \partial_\nu }{2\pi}\begin{pmatrix}
    \alpha^{(\uparrow)}_\eta \\  \alpha^{(\downarrow)}_\eta
   \end{pmatrix}\\
   =& \frac{\epsilon^{\mu\nu\eta} \partial_\nu }{2\pi} \begin{bmatrix}
2 & 1 \\
1 & 2
\end{bmatrix}^{-1} \begin{pmatrix}
    A^{(\uparrow)}_\eta + \omega_\eta - a^{(\uparrow)}_\eta \\  A^{(\downarrow)}_\eta + \omega_\eta - a^{(\downarrow)}_\eta 
   \end{pmatrix}\label{eq: fluxattachment}
\end{align}
with the standard definition of currents:
\begin{align}
    J^{(\sigma),0} & = \rho^{(\sigma)} = f^{(\sigma),\dagger}f^{(\sigma)}  \\
    J^{(\sigma),i} & = \bm{J}^{(\sigma)}_i= \frac{1}{m}\text{Re} \left[f^{(\sigma),\dagger} \mathrm{i} D^{(\sigma),i} f^{(\sigma)} \right]
\end{align}
From these relations, one can readily see that each pseudospin-$\sigma$ CF binds $2$ fluxes of $a^{(\sigma)}$ and $1$ flux of $a^{(-\sigma)}$, as illustrated in Fig.~\ref{fig: schematic}. With equations, this means
\begin{align}
    \frac{\nabla\times \left(\bm{A}-\bm{a}^{(\sigma)}\right)}{2\pi} = 2\rho^{(\sigma)} + \rho^{(-\sigma)}.
\end{align}
Therefore, in the case of balanced fillings of the two layers with $\nu^{(\updownarrow)}=1/3$, $\langle \nabla \times \bm{a}^{(\uparrow)}\rangle = \langle \nabla \times \bm{a}^{(\downarrow)}\rangle =0$, the CFs see no fluxes on average and should form two separate Fermi seas on two layers before considering the interactions.

Technically speaking, $a^{(\sigma)}$ are not ordinary $U(1)$ gauge fields but are \spinc connections (see App.~\ref{app: parton} for detailed discussions), whose elementary source fields are fermions and $\pi$ holonomy on a non-contractible loop corresponds to a change in the associated fermion boundary condition. Although this distinction does not alter any of the responses, it helps track the spin-charge relation in the resulting theories~\cite{SEIBERG2016395,10.1093/ptep/ptw083,SENTHIL20191,han2023fractional} and the correct statistics of the anyons. As a bookkeeping device, we will use Latin/Greek letters in lowercase to refer to \spinc/ordinary $U(1)$ dynamical gauge fields.

\subsection{Pairing assumption}
\label{sec: pairing}

We treat the interaction part of the theory, $\mathcal{L}_\text{int}$, in an effective way. We assume that the interactions of CFs dominantly give rise to inter-layer, pseudospin triplet pairing in the $p$-wave channels~\footnote{We note that in a different context of $\nu_\text{T}=1$, the pairing of CFs in two layers has previously appeared in ~\cite{PhysRevB.61.10267,PhysRevB.95.085135,PhysRevResearch.5.L042022,PhysRevLett.101.176803}. The intra-layer pairing cases, where the resulting topological orders are non-Abelian, are discussed in Ref.~\cite{PhysRevB.82.233301,PhysRevLett.105.216804}.}. To see why these pairing channels are the relevant ones, below we take a quick detour to the wavefunction perspective: the flux attachment in Eq.~\ref{eq: fluxattachment} amounts to a translation between CF and electron wavefunctions (omitting the Gaussian factors and projection onto the lowest LL)
\begin{align}\label{Eq: Wfn}
    \Psi_e = \Psi_\text{CF} \cdot  \prod_{i<j} (z_i-z_j)^2  (w_i-w_j)^2 \prod_{ij} (z_i-w_j) 
\end{align}
where $\{z_i\}$ and $\{w_i\}$ are the sets of complex coordinates of the electrons/CFs in the two layers: $z_i \equiv x^{(\uparrow)}_i -\mathrm{i} y^{(\uparrow)}_i$ and $w_i \equiv x^{(\downarrow)}_i -\mathrm{i} y^{(\downarrow)}_i$. Then we note that a prototype wavefunction of CF with $p_x + \mathrm{i} p_y$ pairing
\begin{align}
    \Psi_\text{CF} =& \det \left(\frac{1}{z_i- w_j}\right),
\end{align} 
which admits a rewriting using the Cauchy determinant formula:
\begin{align}
    \Psi_\text{CF} =& \prod_{i<j} (z_i-z_j) (w_i-w_j)  \prod_{ij} (z_i-w_j)^{-1},
\end{align} 
which, when combined with Eq.~\ref{Eq: Wfn}, precisely gives rise to the wavefunction of $(330)$ state:
\begin{align}
    \Psi_e =  \prod_{i<j} (z_i-z_j)^3 (w_i-w_j)^3 .
\end{align}
Similarly, the $p_x - \mathrm{i} p_y$ pairing CF wavefunction:
\begin{align}
    \Psi_\text{CF} = \det \left(\frac{1}{\bar{z}_i- \bar{w}_j}\right)
\end{align}
yields an electron wavefunction that has the same nodal and phase winding structure as the $(112)$ state wavefunction
\begin{align}
    \Psi_e = {\mathcal J}_0  \prod_{i<j} (z_i-z_j)  (w_i-w_j) \prod_{ij} (z_i-w_j)^2 
\end{align}
where ${\mathcal J}_0 = \prod_{i<j}|z_i-z_j|^2|w_i-w_j|^2\prod_{ij}|z_i-w_j|^{-2}$ is a non-topological Jastrow factor~\cite{wu1993mixed, PhysRevB.53.15845, PhysRevB.95.085135}. It is thus clear that these pairing assumptions are those relevant in our phenomenological considerations. We note that they may also be derivable from a microscopic treatment of the Cooper instability of the two nested (by the $\mathcal{I}\mathcal{X}$ symmetry) Fermi surfaces.

Coming back to our field theory in Eq.~\ref{eq: L}, the interactions are now treated effectively as:
\begin{align}\label{eq: Lint}
\mathcal{L} &=\sum_{\sigma=\updownarrow} \mathcal{L}_{f}^{(\sigma)}    + \mathcal{L}_\text{CS} +\mathcal{L}_\text{pair} + \mathcal{L}_{\Delta} \\
\mathcal{L}_\text{pair} / \sqrt{g}  &=    \frac{1}{2}\sum_{l, \sigma}\left[ \Delta^{(l),*} f^{(-\sigma)} \mathrm{i} \bm{D}^{(\sigma)} \cdot \left(\bm{e}^1+l \mathrm{i} \bm{e}^2\right) f^{(\sigma)} +\text{h.c.} \right]  \label{eq: Lpair}\\
   \mathcal{L}_\Delta &= \mathcal{K} \left[ D^{(+)}\Delta^{(+)} , D^{(-)}\Delta^{(-)} \right] -\mathcal{V}\left[|\Delta^{(+)}|,|\Delta^{(-)}|\right]\label{eq: L Delta} 
\end{align}
where $\Delta^{(l = \pm)}$  are the pair fields of the $p\pm \mathrm{i}p$ channels, $\mathcal{K}$ is the kinetic term for them, 
\begin{align}
    D^{(l)} =  \partial -\mathrm{i} \left(a^{(\uparrow)} + a^{(\downarrow)} + l  \omega\right)
\end{align}
are their covariant derivative operators for $\Delta^{(l)}$, and $\mathcal{V}$ is a potential that only depends on the amplitudes of the pair fields (chirality-mixing terms such as $\Delta^{(+),*} \Delta^{(-)} $ are disallowed by the rotation symmetry). Note that the $\Delta^{(l)}$ field couples to both $a^{(\uparrow)} + a^{(\downarrow)}$ and the $SO(2)$ spin connection $\omega$ since its phase transforms under spatial rotation with relative coefficients $l$. With fixed $\nu_\text{T}=2/3$ and any $\delta\nu$, the pair fields always see no net flux on average $\left\langle \nabla \times \left(\bm{a}^{(\uparrow)}+\bm{a}^{(\downarrow)}\right)\right\rangle=0$, so that the pairing assumption can be made throughout our discussion in this work.

Since there is no time-reversal symmetry in the system, the theory need not be invariant upon exchanging $\Delta^{(\pm)}$, meaning that $\Delta^{(\pm)}$ can have different condensation amplitudes. Then, the parameters in the potential  $\mathcal{V}$ determine the preferred amplitudes of $\Delta^{(\pm)}$, and $\mathcal{K}$ endows quantum fluctuations around the corresponding uniform saddle-point configuration. There are, in principle, many possible orders that could arise in the pair field sector. However, for simplicity and clarity, we will first focus on the possibilities that one or both of $\Delta^{(\pm)}$ are condensed; that is, the cases where $\vec{\Delta}=(\Delta^{(+)},\Delta^{(-)})$ acquires an expectation value before considering gauge fluctuations. 

If only one of $\Delta^{(l=\pm)}$ condenses, it Higgses the gauge field that couples to it; this effect can be effectively captured by introducing hydrodynamic fields $\eta$ and mutual CS terms~\cite{HANSSON2004497}:
\begin{align}\label{eq: L Delta eff 1}
    \mathcal{L}_{\Delta, \text{eff}} = \frac{1}{2\pi}  \left(a^{(\uparrow)}+a^{(\downarrow)}+ l \omega\right) \mathrm{d} \eta .
\end{align}
In the following, it is often convenient to integrate out $\eta$ and re-parametrize
\begin{align}\label{eq: saddle a}
    a^{(\uparrow)} + \frac{l}{2}\omega  = - a^{(\downarrow)} - \frac{l}{2}\omega \equiv a.
\end{align}
Taking this parametrization, we reach an effective CS description:
\begin{align}\label{eq: LCS'}
    \mathcal{L}_\text{CS}' &= \mathcal{L}_\text{CS}+\mathcal{L}_{\Delta,\text{eff}}\\
    &= \sum_\sigma \frac{1}{2\pi} \left[-\sigma a + A^{(\sigma)}+\left( 1 + \frac{l}{2}\right)\omega\right] \mathrm{d}\alpha^{(\sigma)} \nonumber\\
 & \ \ \ \ \ \ - \frac{1}{4\pi}\begin{pmatrix}
\alpha^{(\uparrow)} & \alpha^{(\downarrow)}
\end{pmatrix} \begin{bmatrix}
2 & 1 \\
1 & 2
\end{bmatrix} \begin{pmatrix}
\mathrm{d}\alpha^{(\uparrow)} \\ \mathrm{d}\alpha^{(\downarrow)}
\end{pmatrix}
\end{align}
An equivalent expression, which can be obtained by `integrating in' an auxiliary \spinc field $b$ that decouples 
\begin{align}\label{eq: integrate in}
   & - \frac{1}{4\pi}\left(\alpha^{(\uparrow)}+\alpha^{(\downarrow)}\right)\mathrm{d}\left(\alpha^{(\uparrow)}+\alpha^{(\downarrow)}\right) \nonumber\\
    \rightarrow &\frac{1}{4\pi} b \mathrm{d} b -\frac{1}{2\pi} b\mathrm{d} \left(\alpha^{(\uparrow)}+\alpha^{(\downarrow)}\right) + \Omega_g
 \end{align}
and then integrating out $\alpha^{(\updownarrow)}$, is (see App.~\ref{app: parton} for a more detailed derivation)
\begin{align}\label{eq: LCS' alter}
    \mathcal{L}_\text{CS}' 
 =& \mathcal{L}_{\text{CS,s}}' + \mathcal{L}_{\text{CS,c}}' + \mathcal{L}_{\text{CS,g}}' \\
 \mathcal{L}_{\text{CS,s}}'=& \frac{2 }{4\pi}  a\mathrm{d} a  -\frac{2}{2\pi}A^\text{s} \mathrm{d} a + \frac{2}{4\pi}  A^\text{s}\mathrm{d} A^\text{s}\label{eq: LCSs' alter}  \\
   \mathcal{L}_{\text{CS,c}}' = & 
 \frac{3}{4\pi}  b\mathrm{d} b -\frac{1}{2\pi} \left[2 A^\text{c}+\left(\frac{1}{2}+l\right)\omega\right] \mathrm{d}b  + \frac{2}{4\pi}  A^\text{c}\mathrm{d} A^\text{c} \label{eq: LCSc' alter}\\
   \mathcal{L}_{\text{CS,g}}' =  &   \frac{1+l}{2\pi}A^\text{c}\mathrm{d}\omega + \left(\frac{5}{4} +l \right)\frac{1}{4\pi}\omega\mathrm{d}\omega+3\Omega_g\label{eq: LCSg' alter}
\end{align}
where $\Omega_g$ is a gravitational CS term that originates from the `framing anomaly' of the CS fields~\cite{PhysRevLett.114.016805,PhysRevB.90.014435}. After `integrating out' all the dynamical degrees of freedom, the coefficient of this term is the chiral central charge of the edge of the system~\cite{SEIBERG2016395}, so we will explicitly keep track of this term, even though we are not going present its detailed expression. The rules include: integrating out particles in a Chern insulator state with Chern number $C$ contribute $C\Omega_g$, and integrating out a dynamical gauge field $\gamma$ with a CS term $\frac{k}{4\pi} \gamma\mathrm{d} \gamma $ contribute $-\text{sgn}(k)\Omega_g$. In this representation, it is clear that the pseudospin and charge sectors are effectively separated, and the states of the CFs, $f$, in the current construction only affects the pseudospin properties. This alternative form of $\mathcal{L}_\text{CS}'$ will be useful in many circumstances below.

On the other hand, if both $\Delta^{(\pm)}$ are condensed, two hydrodynamical fields, $\eta^{(\pm)}$, and the corresponding CS terms in Eq.~\ref{eq: L Delta eff 1} should be introduced. Performing unimodular transformation for the hydrodynamic fields $\eta^{\text{r}}\equiv \eta^{(-)}$ (or $\eta^{(+)}$) and $\eta\equiv \eta^{(+)}+\eta^{(-)}$, the effective theory simplifies to
\begin{align}\label{eq: L Delta eff 2}
    \mathcal{L}_{\Delta, \text{eff}} = \frac{1}{2\pi}   \left(a^{(\uparrow)}+a^{(\downarrow)} + \omega\right) \mathrm{d} \eta + \frac{2}{2\pi} \omega \mathrm{d} \eta^{\text{r}}
\end{align}
Comparing with the previous case, we see that the only difference is the addition of a mutual CS term between dynamic hydrodynamic field $\eta^{\text{r}}$ and the $SO(2)$ spin connection $\omega$ with coefficient $2$. This term indicates that the rotational symmetry is broken down to two-fold (i.e. the system become nematic and the residual symmetry is $C_2$) and  all the orbital spins of the monopoles of the emergent gauge fields  (represented by the coefficient of the mixed CS term between any gauge field and $\omega$) should be defined as mod $2$  any gauge field can be arbitrarily shifted by $\eta^{\text{r}}$  (e.g. $\pm 1 \text{ mod } 2=1$ is understood in the above expression). The necessity of the rotation symmetry breaking can also be seen from the fact that the two pure pairing cases have different geometric and geometric-charge (shift) topological responses. On physical ground, this means the system develops nematic order, which has a special axis specified by the relative phase between $\Delta^{(\pm)}$, $\varrho$, which can fluctuate and give rise to gapless neutral phason excitations, or even restore the rotation symmetry by forming certain `vestigial order'~\cite{svistunov2015superfluid, annurev:/content/journals/10.1146/annurev-conmatphys-031218-013200, grinenko2021state} of the two-component condensate, eventually causing an additional topological order. We defer these detailed discussions to App.~\ref{sec: symmetry}. In the below analysis, we will simply symbolically use ``$+ \frac{2}{2\pi} \omega \mathrm{d} \eta^{\text{r}}$'' to represent the effects of rotational symmetry breaking, whenever needed.

\section{Balanced filling: phases and transitions}
\label{sec: undoped}

\begin{figure*}[th!]
    \centering
\includegraphics[width=0.35\linewidth]{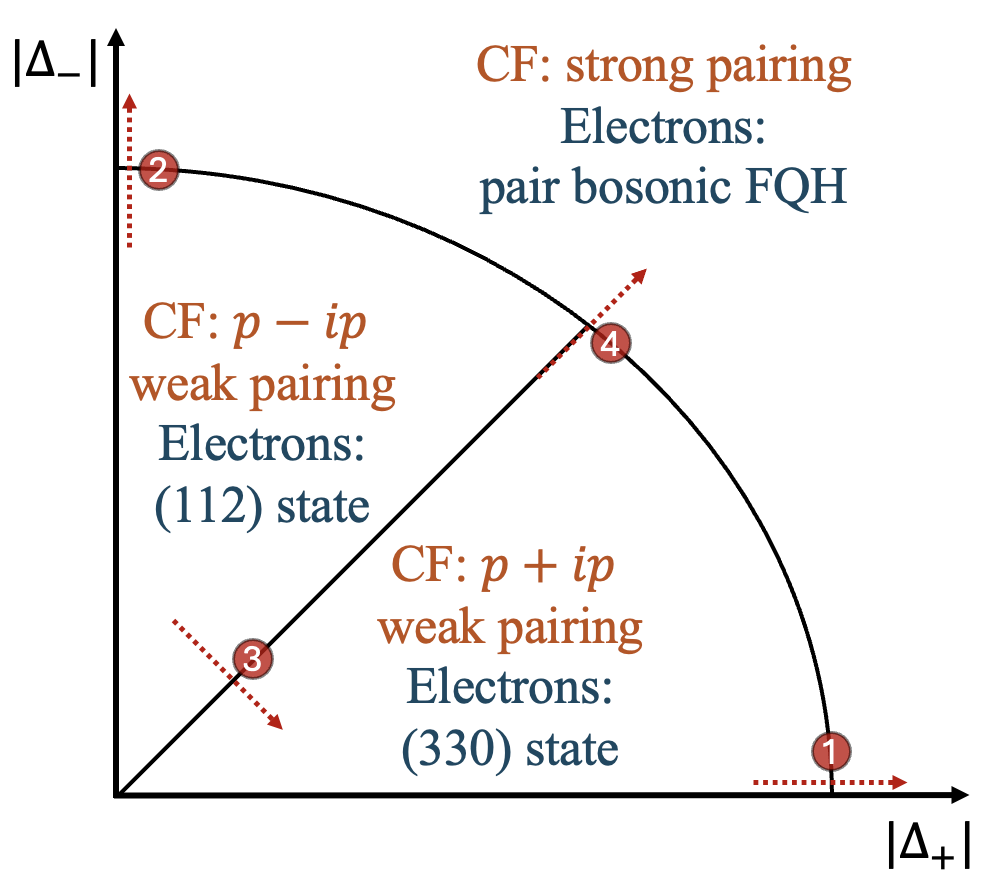}  \includegraphics[width=0.4\linewidth]{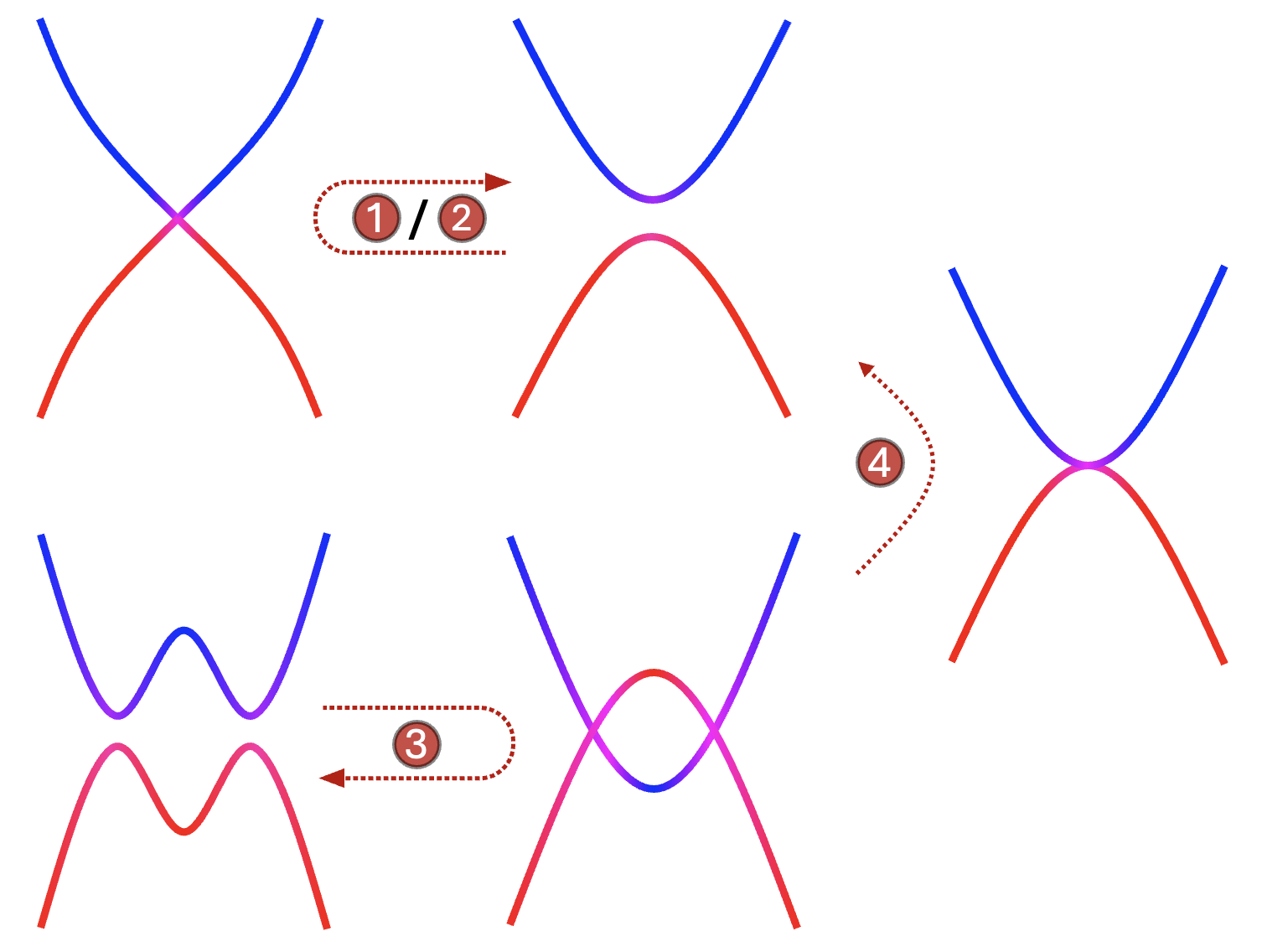}
    \caption{(Left) A mean field phase diagram on $|\Delta^{(+)}|$-$|\Delta^{(-)}|$ plane, valid at the balanced filling $\nu^{(\updownarrow)}=1/3$. The boundary between $p\pm \mathrm{i} p$ phases is specified by $|\Delta^{(+)}|=|\Delta^{(-)}|$, while the boundary of the strong pairing phase depends on details of the model and is set by the condition $\mu=0$ for the fixed filling. For the specific model in Eq.~\ref{eq: BdG}, this boundary is well approximated by $|\vec{\Delta}|\approx 0.13\frac{\bar{B}}{m}$. (Right) Illustrations for the BdG band spectra evolution on different trajectories in the phase diagram. The spectra are plotted along the soft gap/light mass direction in momentum space. Color represents the pseudospin polarization of the eigenmode. Dotted lines mark trajectories across different phase boundaries. }
    \label{fig: CF phase diagram}
\end{figure*}

So far, what remains to be treated in the theory is the paired CFs, which are now more conveniently described using the Nambu basis: 
\begin{align}\label{eq: Nambu CF}
    F \equiv \begin{pmatrix}
        f^{(\uparrow)}\\  - f^{(\downarrow),\dagger}
    \end{pmatrix} . 
\end{align}
Due to the partial charge conjugation, the two components of $F$ couple to $a^{(\uparrow)} = a -  \frac{l}{2}\omega$ and $-a^{(\downarrow)} = a +\frac{l}{2}\omega $, respectively. This means that the fermions $F^{(\uparrow)},F^{(\downarrow)}$ couple to the same gauge field $a$ but have different orbital spins, $- \frac{l}{2}, \frac{l}{2}$, respectively. Hereinafter, we will call $F$ `Nambu CF,' and define the relevant parts  in the total Lagrangian, $\mathcal{L}_f^{(\sigma)}$ in Eq.~\ref{eq: Lf} and $\mathcal{L}_\text{pair}$ in Eq.~\ref{eq: Lpair}, as
\begin{align}\label{eq: L F}
    \mathcal{L}_F \equiv \sum_{\sigma} \mathcal{L}_f^{(\sigma)} + \mathcal{L}_\text{pair}
\end{align}
such that the total effective Lagrangian is simply
\begin{align}\label{eq: L F CS'}
   \mathcal{L}=  \mathcal{L}_F[a]+\mathcal{L}'_\text{CS}.
\end{align}
where we specify the $a$ dependence to highlight that $F$ couple minimally to $a$, and $\mathcal{L}_{\text{CS}}'$ can take either Eq.~\ref{eq: LCS'} or Eq.~\ref{eq: LCS' alter} form.

On flat space, $\vec{\Delta}$ is uniform and $a$ can be taken zero upon gauge transformation. Then the BdG Hamiltonian in the Nambu basis can be simplified in momentum space:  
\begin{align}
    H_F =& \int \frac{\mathrm{d}^2\bm{k}}{(2\pi)^2} F^\dagger_{\bm{k}} h(\bm{k})F_{\bm{k}} \\
    h(\bm{k}) =& \begin{bmatrix}
        \frac{\bm{k}^2}{2m} -\mu  & \sum_{l} \Delta^{(l)} \left( k_x - l \mathrm{i} k_y\right) \\
        \sum_{l} \Delta^{(l),*} \left( k_x + l \mathrm{i} k_y\right) &  -\frac{\bm{k}^2 }{2m} + \mu 
    \end{bmatrix} \label{eq: BdG}
\end{align}
which has an effective particle-hole symmetry. At balanced fillings, the Nambu CFs are at neutrality point, $\rho_F =0$, so the lower band is always fully occupied and the upper band is always empty in the ground state. The value of $\mu$ is determined by the values of $|\Delta^{(\pm)}|$ to ensure the filling fractions. We note that these states provide natural wavefunction ansatzes that could be potentially be compared to numerical results: 
\begin{align}
    \Psi_\text{CF} = \det[\phi(\bm{r}^{(\uparrow)}_i-\bm{r}^{(\downarrow)}_j)]
\end{align}
where $\phi(\bm{r})$ is the Cooper pair wavefunction between $f^{(\updownarrow)}$ obtained by Fourier transforming $u_{\bm{k}}/v_{\bm{k}}$ -- the ratio between the two components of the eigenvector $(u_{\bm{k}},v_{\bm{k}})^T$ to $h(\bm{k})$ in Eq.~\ref{eq: BdG}.

Depending on the condensation amplitudes, $|\Delta^{(\pm)}|$, the CF sector has three possible phases and three phase boundaries; see Fig.~\ref{fig: CF phase diagram} for a summary.  In particular, whenever $\Delta^{(+)}$ and $\Delta^{(-)}$ coexist, the rotation symmetry is broken which manifest as a soft direction in the dispersion of $F$. In the below subsections we establish these results in detail. 

\subsection{$p_x \pm \mathrm{i} p_y$ weak pairing}
\label{sec: p pm p}

We first consider the pure $p_x\pm \mathrm{i} p_y$ weak pairing states with $|\vec{\Delta}|$ small compared to $\frac{\bar{B}}{m}$. In these cases, the BdG spectrum is gapped and the lower band has Chern number $C_F= l = \pm 1$ and orbital spin $s = 0$~\cite{PhysRevX.9.031003,herzog2024interacting}. The corresponding effective theory of $F$ reads:
\begin{align}\label{eq: L Psi eff}
    \mathcal{L}_{F,\text{eff}} =& -  \frac{l}{4\pi} \beta \mathrm{d} \beta + \frac{1}{2\pi} a \mathrm{d} \beta 
\end{align}

Putting together all the pieces of information, we obtain the topological response theory: 
\begin{align}
     \mathcal{L}_\text{top} =&\mathcal{L}_{F,\text{eff}}  + \mathcal{L}'_\text{CS}
\end{align}
For pure $p_x\pm \mathrm{i} p_y$ pairing case, after integrating out $a$, $\beta$ is equated with  $\alpha^{(\uparrow)}-\alpha^{(\downarrow)}$, and we obtain:
\begin{align}\label{eq: L top 1}
     \mathcal{L}_\text{top}
     = & \sum_\sigma \frac{1}{2\pi} \left(A^{(\sigma)}+\omega \pm \omega/2 \right) \mathrm{d}\alpha^{(\sigma)} \nonumber\\
 & \ \ \ \ \ \ - \frac{1}{4\pi}\begin{pmatrix}
\alpha^{(\uparrow)} & \alpha^{(\downarrow)}
\end{pmatrix} \begin{bmatrix}
2 \pm 1 & 1\mp 1 \\
1\mp 1 & 2\pm 1
\end{bmatrix} \begin{pmatrix}
\mathrm{d}\alpha^{(\uparrow)} \\ \mathrm{d}\alpha^{(\downarrow)}
\end{pmatrix} 
\end{align}
which agrees with the Wen-Zee action for Halperin $(330)$ or $(112)$ state~\cite{Wen01101995} and reproduces all the charge, pseudospin, and orbital spin responses, and statistics of the anyons. In particular the shift of the two states (on sphere) are $\mathcal{S}=3,1$ so that they cannot directly transition unless the ratation symmetry is broken as we will discuss below. We thus establish the topological equivalence between these CF $p_x\pm \mathrm{i} p_y$ weak pairing phases and the the Halperin states of electrons in a complimentary way to the wavefunction approach discussed in Sec.~\ref{sec: pairing}.
 
Using the alternative expression for $\mathcal{L}_\text{CS}'$ in Eq.~\ref{eq: LCS' alter}, we can also equivalently write $\mathcal{L}_\text{top}$ as 
\begin{align}\label{eq: L top 2}
    \mathcal{L}_\text{top}
 =& \frac{2 \pm 1 }{4\pi}  a\mathrm{d} a  -\frac{2}{2\pi}A^\text{s} \mathrm{d} a + \frac{2}{4\pi}  A^\text{s}\mathrm{d} A^\text{s}  \pm \Omega_g \nonumber\\
    &  + \mathcal{L}_{\text{CS,c}}' + \mathcal{L}_{\text{CS,g}}' 
\end{align}
Checking the equivalence between Eqs.~\ref{eq: L top 1}\&\ref{eq: L top 2}, one can recognize that the elementary anyons of $a$ or $b$ are the composite anyons of $\alpha^{(\uparrow)},\alpha^{(\downarrow)}$ with composition vector $\vec{l}=(1,-1)$ or $\vec{l}=(1,1)$, respectively~\footnote{When checking the statistical angle, it is important to keep in mind that the $a,b$ are \spinc and thus their anyons' self-statistics should receive an additional $\pi$ phase. It is also important to note that the two alternative representations are not related by a $SL(2,\mathbb{Z})$ transformation but instead are proven equivalent through the procedure in Eq.~\ref{eq: integrate in}. }. These two types of anyons thus can be viewed as the particle-particle pairs and particle-hole pairs of the elementary anyons of $\alpha^{(\updownarrow)}$ on the two layers, respectively. From this expression it is also apparent that the chiral central charges of the two states are $c_-=2,0$ (see discussions below Eq.~\ref{eq: LCS' alter} for the prescription of calculating this quantity).

Since the BdG band is always gapped in the weak pairing regime except when $|\Delta^{(+)}|=|\Delta^{(-)}|$, turning on a weaker coexisting pairing field does not change the BdG band topology from the single-component paring case, so the topological response theory remains basically the same and the only needed change is to add the term $\frac{2}{2\pi} \eta^{\text{r}} \mathrm{d} \omega$ to the $\mathcal{L}_\text{top}$, which represents the physics of rotation symmetry breaking as noted. The corresponding states thus describe nematic versions of the Halperin states. This rotation symmetry breaking can be understood as due to the existence of a soft direction - the special axis specified by the relative phase $\varrho$ between $\Delta^{(\pm)}$ - in the BdG spectrum, which has two gap minima on the line cut along this direction. With the reduced $C_2$ rotation symmetry, a direct transition between $(330)$ and $(112)$ is now possible since their shifts (on sphere) are only well defined module $2$ and in this sense become the same.

\subsection{Strong pairing}
\label{sec: strong}

When the effective attraction among CFs is strong such that $|\vec{\Delta}|$ is substantial compared to $\bar{B}/m$, the chemical potential $\mu$ must be adjusted accordingly to ensure the density constraint $\rho^{(\sigma)} =\nu^{(\sigma)}\frac{\bar{B}}{m} $; eventually such strong pairing will result in a negative $\mu$ which trivializes the band topology of the BdG Hamiltonian. (For the specific model in Eq.~\ref{eq: BdG}, the critical value of $|\vec{\Delta}|$ is approximately $0.13 \frac{\bar{B}}{m}$.) In this case, $\mathcal{L}_F$ yields trivial responses at low energy and the topological response theory is simply $\mathcal{L}'_\text{CS}$. Eq.~\ref{eq: LCS' alter} thus can directly serve as $\mathcal{L}_\text{top}$. Alternatively, one can work with the equivalent expression Eq.~\ref{eq: LCS'}: After integrating out $a$, $\alpha^{(\uparrow)}-\alpha^{(\downarrow)}$ is Higgesed; denoting $\alpha^{(\uparrow)}=\alpha^{(\downarrow)} = \alpha$, Eq.~\ref{eq: LCS' alter} reduces to:
\begin{align}
     \mathcal{L}_\text{top} =&    \frac{1}{2\pi} \left(2 A^{\text{c}}+2\omega \pm \omega \right) \mathrm{d}\alpha- \frac{6}{4\pi}\alpha
\mathrm{d}\alpha 
\end{align}
for pure $p_x\pm \mathrm{i} p_y$ case. 

As before, for coexisting $|\Delta^{(\pm)}|$ cases, we again just need to add a $\frac{2}{2\pi} \eta^{\text{r}} \mathrm{d} \omega$ term, which make the orbital spin of the state, become always $1$ (mod $2$) in the coexisting phase regime.

This state can be thought of as electrons forming inter-layer, charge-$2$ Cooper pairs tightly bounded in real space, each experiencing $6$ external fluxes; these bosons are at an effective filling fraction $\nu_\text{pair} = 1/6$ and forming a bosonic fractional QH state. We note that a similar QH state have been considered at $\nu_\text{T}=2$~\cite{PhysRevB.53.R13275}. Since this state requires the electrons on the two layers to bind, it is probably less likely to be stabilized unless interlayer attractions are introduced.

\subsection{$p_x + \mathrm{i} p_y$ to $p_x - \mathrm{i} p_y$ transition}

\label{sec: transition p pm p}

After establishing the phases of the BdG Hamiltonian Eq.~\ref{eq: BdG}, we now turn to the transitions between these phases. Of particular interest is the possibility of direct transition between weak $p_x \pm \mathrm{i} p_y$ pairing phases of CFs, which can be achieved by two BdG gap closing points at a pair of inversion symmetric momenta $\bm{k}_0$ and $-\bm{k}_0$, as represented by point 3 in Fig.~\ref{fig: CF phase diagram}. This happens when $|\Delta^{(+)}| = |\Delta^{(-)}|$ in the weak pairing regime, where the pairing gap vanishes along an axis in $k$-space whose orientation depends on their relative phase. Where this axis intersects with the $f^{(\sigma)}$ fermi surface, the BdG gap closes and the SC is nodal, resulting in emergent valley degrees of freedom.

The critical theory can be obtained by performing gradient expansion for the Nambu CF field $F$ around the two emergent valleys, which are called $F^{(\xi=\pm)}$ near momenta $\bm{k}_\xi = \xi \bm{k}_0$:
\begin{align}
    \mathcal{L}_\text{Dirac} = \sum_{\xi = \pm 1} \bar{F}^{(\xi)}\left( \mathrm{i} D_0 \gamma^0  - \xi v_\text{F} \mathrm{i} D_\parallel \gamma^1 - \sqrt{2}|\vec{\Delta}| \mathrm{i} D_\perp \gamma^2 \right) F^{(\xi)}
\end{align}
where $\gamma^{0,1,2} \equiv \sigma^2, \mathrm{i}\sigma^1 , \mathrm{i} \sigma^3$  are the gamma matrices satisfying the Clifford algebra $\{\gamma^\mu,\gamma^\nu\} =2 g^{\mu\nu}$, $\bar{F}^{(\xi)} \equiv {F}^{(\xi),\dagger} \gamma^0 $,  $D\equiv \partial -\mathrm{i} a + 1/2 \omega \gamma^0$ is the covariant derivative, $\parallel$ means the direction along $\bm{k}_0$ and $\perp$ represents its normal direction. This is an anisotropic Dirac theory with two flavors of fermions (of opposite parities), both coupling to a dynamical gauge field $a$.

Together with the fluctuating gauge fields, the full effective Lagrangian describing the critical point is a QED$_3$-CS theory:
\begin{align}
    \mathcal{L}_{\text{QED$_3$-CS}} &= \mathcal{L}_\text{Dirac} + \mathcal{L}'_\text{CS}  + \frac{2}{2\pi} \omega \mathrm{d} \eta^{\text{r}} 
\end{align}
with $N_f =2$ flavors of Dirac fermions and a CS term of level $K=-2$ (see Eq.~\ref{eq: LCS' alter}). It is commonly conjectured that many of the QED$_3$-CS theories flow into $(2+1)$-dimensional conformal field theories in the infrared, thus describes continuous quantum phase transitions. For sufficiently small $N_f$ and $|K|$ however, gauge fluctuations may drive the transition first order. We note that similar QED$_3$-CS theories also emerge at low energy in bosonic integer QH transitions~\cite{PhysRevB.87.045129,PhysRevB.89.195143}, bosonic fractional QH to Mott/superfluid transition~\cite{PhysRevB.89.235116} and transitions involving fractional Chern insulators and related phases~\cite{PhysRevX.8.031015,PhysRevLett.80.5409,divic2024anyon}. Recently, numerical calculations using the ``fuzzy sphere'' technique~\cite{zhou2025} have provided compelling evidence for a continuous transition being realized by the $N_f=1$, $K=3/2$ theory.

In the presence of $C_2$ rotation symmetry (in-plane inversion symmetry $\mathcal{I}$), the only allowed mass term of the Dirac theory is
\begin{align}
    m_\text{D}\left(\bar{F}^{(\xi=+)} F^{(\xi=+)} - \bar{F}^{(\xi =-)} F^{(\xi = -)}\right)
\end{align}
which corresponds to tuning $|\Delta^{(+)}|/|\Delta^{(-)}|$ across $1$. This seems to be a topologically trivial mass; however, remember that the two Dirac fermions have opposite parity, so this mass actually changes the Hall response of the Dirac theory by $2$, as expected. We note that this is reminiscent of the scenario in graphene.

\subsection{Weak to strong pairing transition}
\label{sec: transition weak to strong}

The transitions between the weak pairing and the strong pairing phases, represented by points 1 and 2 in Fig.~\ref{fig: CF phase diagram}, are similarly described by QED$_3$-CS theories at low energy. The only difference is that now there is only one flavor of Dirac fermion, i.e. $N_f=1$, obtained by expanding around the $\bm{k}=0$ point.  The parity of the Dirac fermion, or the sign of the $\pi$ Berry phase around the Dirac point, depends on which of $p_x\pm \mathrm{i} p_y$ weak pairing states the system is transitioning out of, so that the effective CS level are $K=5/2,3/2$, respectively.

A particularly interesting scenario is the tricritical point (point 4 in Fig.~\ref{fig: CF phase diagram}). Here, the Dirac velocity along the soft direction vanishes, and the corresponding dispersion becomes quadratic, making the Dirac cone an infinitely anisotropic `semi-Dirac cone~\cite{PhysRevLett.100.236405}' with zero Berry phase. This can be understood by following the transition line $|\Delta^{(+)}|=|\Delta^{(-)}|$ from the weak pairing regime, and recognizing that the semi-Dirac cone is formed by merging two $C_2$ rotation-related Dirac cones with {\it opposite} parities.The resulting low-energy theory is
\begin{align}
    \mathcal{L}_\text{semi-Dirac} = \bar{F}\left( \mathrm{i} D_0 \gamma^0  - \frac{1}{2m} \left[\mathrm{i} D_\parallel \gamma^1\right]^2 - \sqrt{2}|\vec{\Delta}| \mathrm{i} D_\perp \gamma^2 \right) F
\end{align}
which has quadratic dispersion in only one direction and there is no phase winding around the band touching point.  Therefore, this $z=2$ quantum critical point has a {\it distinct} nature from those built with quadratic band touching formed by merged Dirac cones with the same parity and total Berry phase amplitude $2\pi$~\cite{PhysRevLett.103.046811}, a famous example of which in the QH literature is the Haldane-Rezayi state~\cite{PhysRevLett.60.956,PhysRevB.61.10267}.

\section{Doping pseudospin: anyonic exciton superfluidity}

\label{sec: doped}

In this section we consider slightly imbalanced fillings with doped pseudospin ($\delta \nu \ll 1/3$):
\begin{align}
    \nu^{(\sigma)}=1/3+\sigma \cdot \delta \nu  .
\end{align}
that is, $\rho^{(\sigma)}  =\left(\frac{1}{3} + \sigma \delta \nu \right)\frac{\bar{B}}{2\pi} $ or pseudospin density $\rho_\text{s} =\rho^{(\uparrow)} - \rho^{(\downarrow)} = 2\delta \nu\frac{\bar{B}}{2\pi}$. Below we discuss the possible fates of the doped pseudospin.

\subsection{Anyon kinematics}
\label{sec: kinematics}

To quickly see what state the anyons may form and what the state amounts to, we work with the topological response theory of the parent QH states (neglecting geometric terms in this subsection) 
\begin{align} \label{eq:Ltopo}
    \mathcal{L}_\text{top} =&\frac{2+C_F}{4\pi} a\mathrm{d} a  -\frac{2}{2\pi}A^\text{s} \mathrm{d} a + \frac{2}{4\pi}  A^\text{s}\mathrm{d} A^\text{s}   \nonumber\\
    &  + \mathcal{L}_{\text{CS,c}}'+ (3+C_F) \Omega_g
\end{align}
where $C_F =\mp 1, 0$ is the Chern number of $\mathcal{L}_F$ sector for $p_x\pm\mathrm{i}p_y$  weak pairing and strong pairing phases respectively. Since pseudospin is doped, the relevant doped anyons are those coupled to $a$, which we describe with a {\it fermionic} field $\phi^{a}$ that minimally couple to $a$ (remember $a$ is \spinc  so its source must be a fermionic current). From the equation of motion of $a$ in $\mathcal{L}_\text{top}$, we see that:
\begin{equation}
  \frac{1}{2\pi}\langle\nabla\times\bm{a}\rangle = - \rho_{\phi^a}/(2+C_F),  
  \label{Eq:Filling}
\end{equation} 
and since $a$ couples to $A^\text{s}$ through mutual CS term with level $-2$, $-\frac{2}{2\pi}\langle\nabla\times\bm{a}\rangle = \rho_\text{s} = 2\delta\nu \bar{B}$. From these equations, it can be seen that these anyons carry  pseudospin $2/(2+C_F)$, and assuming the doped pseudospin are carried by these anyons, we find from Eq.~\ref{Eq:Filling} that the $\phi^a$ fields are at effective filling fraction $\nu_{\phi^a}=-(2+C_F)$. A natural ansatz state then is for the $\phi^a$ fermions to  form an integer QH state with the corresponding filling fraction. The response of the integer QH state exactly cancels the self CS term of $a$ in Eq.~\ref{eq:Ltopo}, leading to the topological response theory:
\begin{align}\label{eq: L ES}
    \mathcal{L}_\text{ES} =&  -\frac{2}{2\pi}A^\text{s}\mathrm{d} a + \frac{2}{4\pi}  A^\text{s}\mathrm{d} A^\text{s}  + \mathcal{L}_{\text{CS,c}}' +  \Omega_g 
\end{align}
The absence of self CS term of $a$ and the mutual CS term between $a$ and $A^\text{s}$ with level $2$ together indicate that now $A^\text{s}$ is higgsed by a pseudospin-$2$ condensed boson with current $J = \frac{1}{2\pi} \star\mathrm{d}a$, meaning that the pseudospin $U(1)_s$ symmetry is broken down to $\mathbb{Z}_2$ by the formation of a pseudospin-$2$ ES, which features vanishing longitudinal and Hall resistivity in the pseudospin channel measurable in suitable counter-flow measurements. The residual topological order is purely in charge sector and is described by $\mathcal{L}_{\text{CS,c}}'$, which has the minimal quantum dimension $3$ at this charge filling fraction~\cite{musser2024fractionalization,jian2024minimal}. From the action it can also be read out that the ES state has total chiral central charge $c_- =0$ (see discussions below Eq.~\ref{eq: LCS' alter} for the prescription of calculating this quantity).  

This kinematics is in direct analog to that of anyonic superconductivity. However, this analysis only shows that such a state can exist; whether it can be stabilized against other possible states, such as various sorts of crystalline orders~\cite{PhysRevLett.119.067002,kim2021quantum,PhysRevB.85.241307}, localized states~\cite{nosov2025plateau,shi2025anyon}, and phase separation to other QH states at different filling fractions, remains undetermined since the preceding topological analysis omits details of energetics.  Similarly, since the band structure and the orbital spin of the anyons are not known, we can not determine the angular momentum and geometric responses of the ES constructed above, either. Luckily, we note that the fermionic field $\phi^a$ that was used to describe the anyons is nothing but the doped part of the Nambu CF $F$ particles on top of the filled band. As we will show below, taking one step back and work with the theory of $F$ near criticality yields more information and justification of these anyonic ES states.

\begin{figure}
    \centering
    \includegraphics[width=\linewidth]{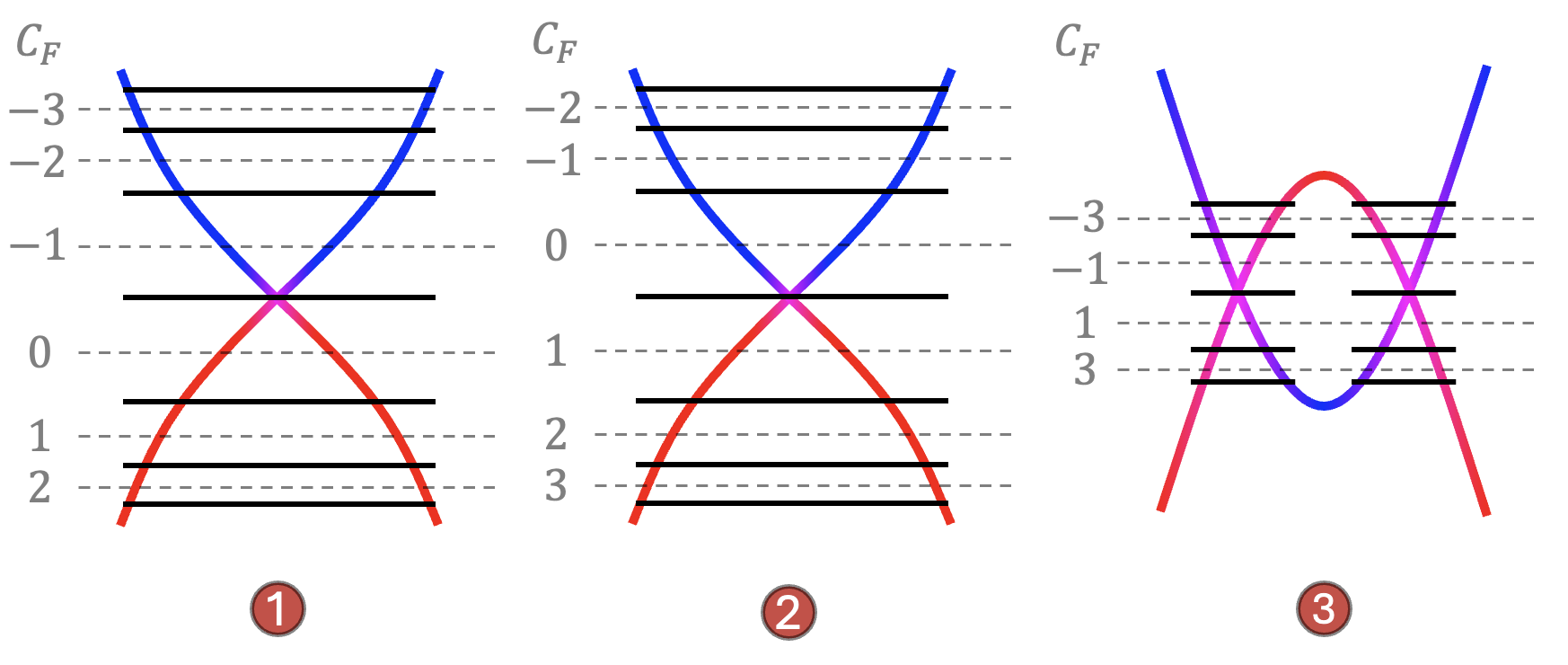}
    \caption{Illustrations of the LL structure and the gap Chern numbers ($C_F$) at the transition points 1/2/3 in Fig.~\ref{fig: CF phase diagram}, corresponding to $p_x + \mathrm{i}p_y$ to strong pairing, $p_x - \mathrm{i}p_y$ to strong pairing and $p_x - \mathrm{i}p_y$ to $p_x + \mathrm{i}p_y$ transitions, respectively. LLs are always filled up to $C_F=-2$ on doping pseudospin. Combining with the fact that the $n$-LL particles carries orbital spin $-n$~\cite{golkar2014effective}, this leads to different angular momentum for the resulting ES: $-1$, $-3$, and $1 \text{ mod } 2$, respectively. This figure assumes $\delta \nu>0$; the $\delta\nu<0$ case can be obtained by flipping the signs of the Chern number and the orbital spin.  }
    \label{fig: Dirac LL}
\end{figure}

\subsection{Composite fermion approach}
\label{sec: mechanism}

The doped anyons can be interpreted as excessive Nambu CF $F$ particles, therefore it is more complete to directly consider the theory of $F$ for the fate of doped pseudospin. Formally, our theory of $F$ is (writing out  Eq.~\ref{eq: L F CS'}): 
\begin{align}\label{eq: L F simple}
\mathcal{L}=\mathcal{L}_F[a]+ \frac{2 }{4\pi}  a\mathrm{d} a  -\frac{2}{2\pi}A^\text{s} \mathrm{d} a + \dots
\end{align}
which largely resembles Laughlin's theory of anyonic superconductivity~\cite{doi:10.1126/science.242.4878.525,PhysRevLett.60.2677,PhysRevB.39.9679}, albeit realized in a totally different setup and with the crucial difference that $F$ has two bands.  

Upon doping pseudospin, $F$ particles see an average flux density $\langle \nabla \times \bm{a}\rangle = -\delta \nu  \cdot \bar{B}$, which rearranges the dispersion of the BdG band in Eq.~\ref{eq: BdG} into LLs. Since $F$ particles have net density $\rho_{F} = \rho_\text{s} = 2\delta\nu \cdot \bar{B}/(2\pi)$, they are at effective filling fraction $\nu_F = -2$, and thus a natural ansatz state is to occupy an appropriate number of LLs to yield $C_F=-2$, which is represented by $\mathcal{L}_F=\frac{-2}{4\pi} a \mathrm{d} a$. This immediately yields the same topological response theory as Eq.~\ref{eq: L ES} obtained in the last section from the anyon kinematics. 

We now further analyze the properties of this ES state near different transition lines in Fig.~\ref{fig: CF phase diagram}. By tracking the LL gaps of the Dirac fermions across the transition and matching the one that will evolve into the band gap in the zero field limit, we are able to explicitly obtain the Chern number of each LL gap near each type of transition, as shown in Fig.~\ref{fig: Dirac LL}. Moreover, Ref.~\cite{golkar2014effective} has computed the orbital spin of $n$-th LL of a Dirac fermion to be $-\text{sgn}(\delta\nu)n$ (the sign depends on the direction of the effective magnetic field; we note that the result is different from the non-relativistic case~\cite{PhysRevLett.69.953}). Taking these into account, we obtain a refined version of topological response theory, $\mathcal{L}_\text{ES}$, for ES near the transition points in Fig.~\ref{fig: CF phase diagram}.

First let us consider the cases of CF $p_x \pm \mathrm{i}p_y$ to strong pairing transitions ($(330)$/$(112)$ to bosonic FQH transitions represented by points 1 and 2 in Fig.~\ref{fig: CF phase diagram}). The simplification here is that the presence of a single Dirac point at Gamma allows us to preserve spatial symmetries. The resulting theory is
\begin{align}\label{eq: L ES refined}
    \mathcal{L}_\text{ES} =&  -\frac{1}{2\pi}\left[2A^\text{s} -(2\pm 1)\omega\right]\mathrm{d} a + \frac{2}{4\pi}  A^\text{s}\mathrm{d} A^\text{s}  -\frac{3\pm 2}{4\pi} \omega\mathrm{d}\omega  \nonumber\\
    & + \mathcal{L}_{\text{CS,c}}' + \mathcal{L}_{\text{CS,g}}' -2 \Omega_g
\end{align}
which implies that the angular momentum of the ES is $-3$ or $-1$, depending on whether one is emerging from the $(330)$ or the $(112)$ state respectively. Apart from this fact, the ES emerging from both these states have the same background topological order (of the $(112)$ type) and the same edge central charge $c_-=0$. This fact indicates that these two ES states cannot be connected without breaking the rotation symmetry.

Pseudospin doping of the system in the general vicinity of the $p_x +\mathrm{i}p_y$ to  $p_x - \mathrm{i}p_y$ weak pairing transition ($(112)$ to $(330)$ transition) represented by point 3 in Fig.~\ref{fig: CF phase diagram} is slightly more subtle: there are two degenerate LLs (belonging to the two emergent valleys) at the Fermi level, and the filling fraction $\nu_F=-2$ corresponds to only one of them being occupied. Note that although we did not explicitly include the interactions among $F$ particles, there should generically be residual interactions that drive QH ferromagnetism splitting the LLs, which leads to additional weak (i.e. amplitude proportional to $|\delta\nu|$) symmetry breaking. Depending on the details of the interaction, the possible valley orders include e.g. valley polarization or inter-valley coherence order. In terms of the physical observables, the first possibility leads to further breaking of the remnant $C_2$ symmetry, and the second possibility leads to translation symmetry breaking at wavevector $2\bm{k}_0$ and a charge density wave for $F$.  In the second case, the physical pattern of the electrons is a pseudospin density wave (according to the EOM of Eq.~\ref{eq: L F CS'}), and it still makes sense to compute the angular momentum of the ES, which we find to be $1$ mod $2$.

The most important aspect of this CF analysis, which is also a central result of this work, is an estimation of the energy of the ES state near the critical points. For clarity, we assume $\delta \nu >0$ in the below discussion without loss of generality. Assuming the $\Delta$ sector is not singularly perturbed by the light doping (i.e. the relevant energies are $\mathcal{O}(\delta\nu^2)$), the leading energy cost comes from $F$ occupying extra LLs above charge neutral point of the Dirac cone. As a function of $\delta\nu$, we compute this energy per Dirac cone to be:
\begin{align}
    E(\delta \nu) =& A \frac{v_\text{F}\sqrt{2 \bar{B}^3}}{2\pi}\left[ -|\delta\nu|^{3/2}\zeta\left(-1/2,\frac{\tilde{m}_\text{D}^2}{|\delta\nu|}+n_0+1\right)\right.\nonumber\\
    &\ \ \ \ \ \ \ \ \ \ \ \ \ \ \ \ \ \ \ \left. + \frac{1}{2}|\delta\nu |\tilde{m}_\text{D} - \frac{2}{3}|\tilde{m}_\text{D}|^3 \right] + \mathcal{O}(\delta\nu^2)
\end{align}
where $A$ is the area of the sample, $\zeta$ is the Hurwitz zeta function, $v_\text{F}$ is the Fermi velocity of the Dirac cone~\footnote{For anisotropic Dirac cones, this should be the geometric mean of the Dirac velocities along two directions.}, $\tilde{m}_\text{D}\equiv m_\text{D}/(v_\text{F}\sqrt{2\bar{B}})$ is a dimensionless parameter for characterizing the Dirac mass $m_\text{D}$ (the coefficient of $\xi \bar{F}F$ in the Lagrangian density where $\xi$ is the parity of the Dirac cone), and $n_0$ is the number of occupied LLs above the zeroth LL. $n_0=1,2$ for transitions 1,2 in Fig.~\ref{fig: CF phase diagram}, and $n_0 =0,1$ for the two Dirac cones emerging at the transition 3 in Fig.~\ref{fig: CF phase diagram}. This is a singular function where $\delta\nu \rightarrow 0$ and $\tilde{m}_\text{D}\rightarrow 0$ limits do not commute. 

When $\tilde{m}_\text{D}\neq 0$,
\begin{align}
    E(\delta \nu) \approx  A \frac{v_\text{F}\sqrt{2 \bar{B}^3}}{2\pi}\left[\frac{1}{2}\tilde{m}_\text{D}+\left(n_0+\frac{1}{2}\right)|\tilde{m}_\text{D}|\right] |\delta\nu| + \mathcal{O}(\delta\nu^2)
\end{align}
which indicates a finite pseudospin gap at zero doping
\begin{align}
    \Delta_\text{s} \equiv \frac{\pi}{A\bar{B}} \left[\left.\frac{\partial E}{\partial \delta\nu}\right|_{\delta\nu\rightarrow 0+} - \left.\frac{\partial E}{\partial \delta\nu}\right|_{\delta\nu\rightarrow 0-}\right]
\end{align}
of the system, as expected. On the other hand, when $\tilde{m}_\text{D} = 0$, the LL level structure of the Dirac cone yields
\begin{align}
    E(\delta \nu) \approx \zeta (-1/2,n_0 +1) A \frac{v_\text{F}\sqrt{2 \bar{B}^3}}{2\pi} |\delta\nu|^{3/2} + \mathcal{O}(\delta\nu^2)
\end{align}
which indicates that the pseudospin gap vanishes. This scaling is further softened to $|\delta\nu|^{5/3}$~\cite{PhysRevLett.100.236405} at the semi-Dirac tricritical point, i.e. the transition 4 in Fig.~\ref{fig: CF phase diagram}. 

The fact that the leading order energy is as soft as $|\delta\nu|^{3/2}$ when doping the critical point can be understood as a consequence of the emergent Lorentz invariance at the critical point described by QED$_3$-CS theories, where time and space have the same scaling and characteristic length scale is $\sim 1/\sqrt{\delta\nu\bar{B}}$. This argument strongly suggests that this ES state may have lower energy than other competing states, such as phase separation to other QH states, which generically do not have Lorentz invariance and thus typically have a harder energy scaling $E\sim |\delta\nu|$. We note that the weak splitting of LLs in the case of doping $(330)-(112)$ transition only cause a subleading $\mathcal{O}(\delta\rho^2)$ contribution, thus does not affect this scaling in the dilute limit. The power in the energy scaling $E\sim |\delta\nu|^\eta$ can in principle be measured capacitively~\cite{PhysRevB.84.085441,zibrov2017tunable} through the characteristic layer polarization ($P\propto \delta\nu $) dependence on the perpendicular displacement field ($D_z\propto \mu^{(\uparrow)}-\mu^{(\downarrow)}$): $P \sim D_z^{1/(\eta-1)}$. For $\eta \neq 1$ this leads to a power-law polarizability whereas $\eta=1$ corresponds to an `incompressible' state of pseudospin with a finite spin gap as computed above. 

Within this FP mean field picture, we are also able to estimate the superfluid stiffness $\kappa_\text{s}$ of the ES (see App.~\ref{app: sec: stiffness} for details). We find that the scaling behavior depends similarly on whether the $F$ bands are gapped:
\begin{align}
    \kappa_\text{s} \sim \begin{cases}
        |\delta\nu|^{1/2} & \tilde{m}_\text{D}=0\\
        |\delta\nu| & \tilde{m}_\text{D} \neq 0
    \end{cases}
\end{align}
This is a striking result suggesting that the ES obtained upon doping the critical point has an anomalously high stiffness that can quickly stabilize this phase at low temperature, since the Berezinskii–Kosterlitz–Thouless transition temperature $T_c$ of superfluidity is usually determined by the stiffness. It is also interesting to note that the scaling of the number compressibility and the stiffness together suggests that the speed of sound of the superfluid approaches a constant as $\delta\nu\rightarrow 0$. To the best of our knowledge, we are not aware of other mechanism for superfluidity that gives rise to the same scaling behavior of the stiffness, so we propose this phenomenon as the distinguishing feature of the mechanism based on anyons with Dirac dispersions. { We note that this scaling of stiffness is not affected by the subtlety associated with the case of doping $(330)-(112)$ transition where a weak splitting of LLs occurs (App.~\ref{app: sec: stiffness}).}

{ We remark that the above energetic considerations as well as the discussions of the anyon dispersion are derived within a mean field theory and may receive further corrections from gauge fluctuation effects. Nonetheless, we note that these results can be derived more generally from the assumed 
conformal invariance of the critical point through a scaling argument~\cite{monin2017semiclassics}. If the infrared fate of the theory is actually not conformal (for example, if the transition is weakly first order or flows to a different dynamical exponent), then the scaling relations need not strictly apply and the energetics could be modified accordingly. }

In real-world experiments, one must contend with disorder, which usually tend to stabilize the gapped QH state - leading to QH plateaus in transport measurements - when quasi-particles are doped. The combined roles of disorder and interactions are of course difficult to model. However, we can nonetheless obtain some qualitative results by assuming the $F$ particles are non-interacting fermions at low energy, whose properties under a disorder potential are well studied. Within this picture, the effective LLs are broadened, most of the states except one that carries the Hall response will be localized, and the only extended state within each LL will be levitated in energy on increasing disorder strength. All of these facts will result in a non-zero critical pseudospin density at which the ES state onset, even without any competition from other states~\cite{nosov2025plateau,shi2025anyon}. In App.~\ref{app: disorder} we use a simple model to describe this physics, and we find 
\begin{align}
    |\delta\nu_{\text{critical}}| \propto \tilde{m}_\text{D}^2
\end{align}
for sufficiently weak disorder. The plateau width scales with the square of the Dirac mass, reflecting the fact that massless Dirac fermions are intrinsically harder to localize. and  again highlighting the stability of ES when doping the Dirac point.

\section{Generalizations}
\label{sec: generalization}

\subsection{Bilayer Jain sequence: a direct generalization}
\label{sec: principal}

As discussed in App.~\ref{app: parton}, we find that the CF theory adopted in this work can be readily generalized to other filling fractions -- especially those in the principal Jain sequence, $\nu^{(\updownarrow)}=n/(2n+1)$ with $n$ an arbitrary nonzero integer. Indeed, a recent experiment~\cite{nguyen2025bilayerexcitonslaughlinfractional} has identified evidence for a transition out of the decoupled bilayer QH state, upon varying $d/\ell_B$, not only at filling fractions $\nu^{(\updownarrow)}=1/3$ but also $2/5$ and $3/7$. An earlier experiment~\cite{liu2019interlayer} indeed has found transport signatures that cannot be explained by decoupled QH states at a relatively small $d/\ell_B$ at $\nu^{(\updownarrow)}=2/5$ and $3/7$. In this section, we briefly discuss the scenario predicted by the simplest generalization of our theory for the transition at these filling fractions corresponding to the principal Jain sequence. 

In the generic cases with $n\neq 1$, the $f$ fields no longer admit the interpretation as a CF, but should rather be viewed as a fermionic parton (FP). As shown in App.~\ref{app: parton}, the flux attachment scheme of the CF theory in Eq.~\ref{eq: fluxattachment} should be generalized to:
\begin{align}
     \mathcal{L}_\text{CS}  =&  \sum_{\sigma,I=1,\dots,n}  \frac{1}{2\pi} \left(A^{(\sigma)}-a^{(\sigma)} + I \omega \right) \mathrm{d}\alpha^{(\sigma,I)} \nonumber\\
     &\ \ \ \ \ - \frac{1}{4\pi}
\vec{\alpha}^{T}  ({\bf I }+  {\bf E })_{2n\times 2n}  \mathrm{d}
  \vec{\alpha}\nonumber\\
 \vec{\alpha} \equiv & \begin{pmatrix}
\alpha^{(\uparrow,1)} & \dots & \alpha^{(\uparrow,n)}  & \alpha^{(\downarrow,1)} & \dots & \alpha^{(\downarrow,n)}
\end{pmatrix}^T
\end{align}
where ${\bf I}$ is the identity matrix and ${\bf E}$ represents a matrix with all entries equal to $1$. It is straightforward to check that when $n=1$, the action reduces to the CF theory in Eq.~\ref{eq: fluxattachment}; whereas when $n\neq 1$, the attached intralayer flux to each $f$ particle will be fractional, rendering the CF interpretation invalid, so that $f$ must be understood as FP.

Then, in the analysis for $\Delta$ and $f$ that precisely parallels the previous case, we find the weak pairing $p_x\pm\mathrm{i} p_y$ phases of the $f$ FPs corresponds to QH states characterized by a $2n\times 2n$ $K$-matrix for emergent $U(1)$ gauge fields $\vec{\alpha}$:
\begin{align}
K &= \left[\begin{array}{ c | c }
   \left[{\bf I} +(1\pm 1) {\bf E}\right]_{n\times n} & (1\mp 1) {\bf E}_{n\times n} \\
   \hline
  (1\mp 1) {\bf E}_{n\times n} & \left[{\bf I} + (1\pm 1) {\bf E}\right]_{n\times n}
  \end{array}\right] 
  \label{Eq:JainHierarchy}
\end{align}
with charge vector $c=(1,\dots,1,1,\dots,1)^T$, pseudospin vector $p=(1,\dots,1,-1,\dots,-1)^T$ and orbital spin vector $s = (1\pm 1/2,\dots, n\pm 1/2,1\pm 1/2, \dots, n\pm 1/2)^T$. One can check that the weak $p+\mathrm{i} p $ pairing which should have $C_F =1$ and $L=1$ indeed gives rise to the topological order of a pair of decoupled Jain states~\cite{Wen01101995}. We will comment further on the weak pairing $p-\mathrm{i} p $ phase below. The strong pairing phase, on the other hand, has a $K$-matrix:
\begin{align}
K &= \left[\begin{array}{ c | c }
   \left[{\bf I} + {\bf E}\right]_{n\times n} &  {\bf E}_{n\times n} \\
   \hline
  {\bf E}_{n\times n} & \left[{\bf I} + {\bf E}\right]_{n\times n}
  \end{array}\right] 
\end{align}
subject to an additional constraint 
\begin{align}\label{eq: additional constraint}
    \sum_{I} \alpha^{(\uparrow,I)} = \sum_{I} \alpha^{(\downarrow,I)}.
\end{align}

A more straight-forward account for the topological  properties can be achieved with the spin-charge separated flux attachment scheme similar to Eq.~\ref{eq: LCS' alter} (omitting geometric terms here; full expression can be found in App.~\ref{app: parton}):
\begin{align}\label{eq: LCS' alter_generalized}
\mathcal{L}=& \mathcal{L}_F[a] + \mathcal{L}_{\text{CS,s}}' + \mathcal{L}_{\text{CS,c}}' +\dots \\
 \mathcal{L}_{\text{CS,s}}'=& \frac{2 n }{4\pi}  a\mathrm{d} a  -\frac{2 n }{2\pi}A^\text{s} \mathrm{d} a + \frac{2 n}{4\pi}  A^\text{s}\mathrm{d} A^\text{s} \\
   \mathcal{L}_{\text{CS,c}}' = & 
 \frac{2n+1}{4\pi}  b\mathrm{d} b -\frac{2n}{2\pi} A^\text{c}\mathrm{d}b  + \frac{2n}{4\pi}  A^\text{c}\mathrm{d} A^\text{c} 
\end{align}
For the weak pairing $p_x \pm \mathrm{i}p_y$ or strong pairing phases, $\mathcal{L}_{F}[a] = \frac{C_F}{4\pi} a\mathrm{d} a$ where $C_F=\pm 1, 0$ is the Chern number of $F$ band. The relevant neutral anyons described by $F$, which minimally couple to $a$, are the anyonic dipoles formed by a pair of $nC_F/(2n+C_F)$-charge anyons across the two layer; in particular, for the decoupled Jain state, $C_F=1$, and the relevant anyon is a bound state formed by $n$ elementary anyonic dipoles each with charges $(\frac{1}{2n+1},-\frac{1}{2n+1})$. Clearly, this mechanism requires a particular energetic hierarchy of the anyon dipoles, which is naturally achieved near the gap closing transition of $F$.

Within the effective description Eq.~\ref{eq: LCS' alter_generalized}, it can be readily seen that our discussion about the criticality and the consequence of pseudospin doping can be transferred with few modifications. A continuous transition between $p\pm \mathrm{i}p$ weak pairing phases again  necessarily implies a rotation symmetry breaking. Assuming such a symmetry reduction is present, the critical point at balanced filling, which is accessed by tuning the interlayer correlations, is predicted to be QED$_3$-CS with $N_f=2, \,K=-2n$.

Upon pseudospin-doping this critical point, the Nambu FP $F$ which couples to $a$ are at effective filling fraction $\nu_F = -2n$ according to the self CS level of $a$. Filling an appropriate number of LLs in the $F$ spectrum again cancels the self CS term $\frac{2n}{4\pi} a\mathrm{d} a$ and result in a ES that breaks the pseudospin $U(1)$ symmetry to $2n$ fold, i.e. the elementary particle in the ES condensate has pseudospin $2n$. A more systematic characterization of the properties of all these phases and transitions will be given in the next subsection.

For concreteness, let's walk through the particular example of $\nu^{(\updownarrow)}=2/5$ in each layer corresponding to $n=2$. Based on our theory, we expect a quantum phase transition between two states anticipated at small and large $d/\ell_B$, which are specified by the $K$ matrices: 
\begin{align}
K_{d \lesssim \ell_B} = \begin{bmatrix}
1 & 0 & 2& 2 \\
0 &  1 & 2& 2 \\
2 & 2& 1& 0 \\
2 & 2& 0& 1
\end{bmatrix}; \ \ K_{d\gg \ell_B} = \begin{bmatrix}
 K_{2/5} & 0 \\
0 & K_{2/5}
\end{bmatrix}
\end{align}
where $K_{2/5}=\begin{bmatrix}
3 & 2 \\
2 & 3
\end{bmatrix}$ is the K-matrix of the $\nu=2/5$ single-layer Jain state. The charge vector is $c=(1,1,1,1)^T$, pseudospin vector is $p=(1,1,-1,-1)^T$ and orbital spin vector is $s=(1\mp 1/2, 2\mp 1/2,1\mp 1/2, 2\mp 1/2)^T$. From these we can read off that the ground state degeneracies are $5\times 3,\, 5\times 5$ and the pseudospin Hall conductances, which  are $-4/3, 4/5$, and the shifts, are $2, 4$ for the two states characterized by $K_{d \lesssim \ell_B}$ and $K_{d  \gg_B}$, respectively. 

This transition can be understood from a complimentary perspective. The relevant, softened anyonic exciton excitation on the large $d/\ell_B$ side are charge $(2/5,\,-2/5)$ dipoles. This can be described by a bosonic field that couple to the emergent gauge fields $\vec{\alpha}$ with composition vector $l=(1,\,1,\,-1,\,-1)^T$. When this is simply gapped we are in the decoupled Jain state, while when it undergoes a transition into a bosonic integer QH phase, it leads to the other phase. This bosonic topological order leads to change in the $K$ matrix:  \begin{align} 
\Delta K = \begin{bmatrix}
-2 & -2 & 2& 2 \\
-2 &  -2 & 2& 2 \\
2 & 2& -2& -2 \\
2 & 2& -2& -2
\end{bmatrix}; K_{d \lesssim \ell_B} = K_{d \gg \ell_B} + \Delta K
\label{Eq:KforJain25}
\end{align}
and stronger inter-layer correlations. At the wavefunction level, this is captured by the Jastrow-type factor desribing a bosonic IQH state: 
\begin{align}
    \mathcal{J}(z,w) = \prod_{ij}(z_i-w_j)^2\prod_{i<j}(z_i-z_j)^{-2}\prod_{i<j}(w_i-w_j)^{-2},
\end{align}
where $z,\,w$ refer to complex coordinates of electrons in the top and bottom layers. When it is attached to the large-$d/\ell_B$ wavefunction, we get the small-$d/\ell_B$ wavefunction:
\begin{align}
     \Psi_{d \lesssim l_B}(z,w) = \mathcal{J}(z,w) \Psi_{d \gg l_B}(z,w)
\end{align}
which clearly has stronger inter-layer electron avoidance that can help reducing the inter-layer Coulomb energy.

Since the low-energy anyonic excitons associated with this transition are charge $(2/5,\,-2/5)$ dipoles, the minimal {\it local} operator built out of them is an charge $(2,-2)$ bosonic excitation, instead of the elementary dipole composed of an electron hole pair $(1,-1)$. The condensation of these objects can therefore lead to a pseudospin-$4$ exciton condensation, which does not manifest in any fermion bilinear observables of the system. This is an exotic and long-sought-for spin nematic order, and we predict the realization of it when doping the critical point at higher Jain fractions.

\subsection{Bilayer Jain sequence: a further generalization}
\label{sec: further}

Here we make a further generalization to our theory by allowing the pairing of $f$ in arbitrary pairing channels along with arbitrary BdG band structure. This is readily achieved by using the theory in Eq.~\ref{eq: LCS' alter_generalized}, based on which we present some general results in the following.

If the angular momentum of the pairing $L$ is known, the shift (on sphere) of the resulting state is (App.~\ref{app: parton})
\begin{align}
    \mathcal{S} = n+1+L
\end{align}
[We note that, when the rotation symmetry is broken down to $C_m$, then only the mod-$m$ part of the shift is well defined.] Similarly, if $F$ band is gapped and the Chern number of the occupied band $C_F$ is known, the total chiral central charge is then (defining $\text{sgn}(x)=x/|x| $ for $x \neq 0$ and $\text{sgn}(0)= 0 $)
\begin{align}
    c_- = 2n + C_F - \text{sgn}(2n+C_F).
\end{align}
This case also implies a well defined topological degeneracy $g= (2n+1)\times |2n+C_F|$ ($C_F \neq -2n$) where the two factors are the degeneracy in the charge and pseudospin sectors.

Based on the Ioffe-Larkin composition rule, one can characterize the transport properties of the system with the resistivity tensor of the state formed by the $F$, which we call $\hat{R}^F$:
\begin{align}
 \hat{R} =&  \begin{pmatrix}
     (1+\frac{1}{n})\hat{R}^0+ \hat{R}^F & \hat{R}^0-\hat{R}^F \\
     \hat{R}^0 -\hat{R}^F &(1+\frac{1}{n})\hat{R}^0+ \hat{R}^F
 \end{pmatrix}
\end{align}
where the four channels of the bilayer resistivitity tensor $\hat{R}$ are $(\uparrow x, \uparrow y, \downarrow x, \downarrow y)$ and $\hat{R}^0 \equiv  2\pi\begin{pmatrix}
         0 & -1 \\
         1 & 0 
\end{pmatrix}$.

When $F$ is gapped and $C_F\neq 0, -2n$,  we have $ \hat{R}^F= \frac{1}{C_F} \hat{R}^0$. This amounts to intra- and inter-layer resistivity (commonly called `drive' and `drag' in experiments)
\begin{align}
    R^\text{drive}_{xx} =& R^\text{drag}_{xx} = 0 \ , \\
     \frac{1}{2\pi}R^\text{drive}_{yx} =& 1+\frac{1}{n} +\frac{1}{C_F} \\
     \frac{1}{2\pi}R^\text{drag}_{yx} =
     &1-\frac{1}{C_F}  
\end{align}
or in terms of conductivity in charge and pseudospin channels, vanishing longitudinal conductivity $\sigma^{\text{c}}_{xx} = \sigma^{\text{s}}_{xx}=0$ and fractional Hall conductivity $2\pi \sigma^{\text{c}}_{xy}=\frac{2n}{2n+1}$ and $2\pi \sigma^{\text{s}}_{xy}=\frac{2n C_F}{2n+C_F}$. A special care needs to be paid to the case  $C_F=0$: in this case the $\hat{R}^{F}$ will be that of a insulating state $\sim (\infty) \mathbf{I}_{2\times 2}$, yielding divergent $R^\text{drive}_{xx}$ and $-R^\text{drag}_{xx}$, but finite $\frac{1}{2\pi} R^\text{drive}_{yx} = 1+1/n$ and $\frac{1}{2\pi} R^\text{drag}_{yx} = 1$.

If $C_F=-2n$, either due to pseudospin doping and filling or a special band topology of $F$, the resulting state would be an ES with a residual topological order purely in the charge sector with anyon count $2n+1$ (still minimal at the fraction) and zero chiral central charge. In this case, the above Ioffe-Larkin rule results are invalid, and the transport properties should instead be characterized by divergent $\sigma^\text{s}_{xx}$ and vanishing $\sigma^\text{s}_{xy}$ with unchanged responses in the charge sector. These amounts to the following resistivity:
\begin{align}
    R^\text{drive}_{xx} =& R^\text{drag}_{xx} = 0 \ , \\
     \frac{1}{2\pi}R^\text{drive}_{yx}  = &
     \frac{1}{2\pi}R^\text{drag}_{yx} =
1+\frac{1}{2n}
\end{align}

Similar to the transition between $p\pm\mathrm{i} p$ phases, which have angular momentum $L=\pm 1$ differing by $2$, any direct transition between two pairing states with two different angular momenta $L_{1,2}$ must at least break rotation symmetry to $|L_1-L_2|$-fold. For the transition between two phases with different Chern number $C_{F;1,2}$, the critical theory generically is a QED$_3$-CS theory with Dirac cone flavor $N_f = |C_{F,1}-C_{F,2}|$ and CS level $K= -(4n + C_{F,1}+C_{F,2})/2$.

All our discussions in Sec.~\ref{sec: mechanism} about the role of disorders can be carried to the current case. 
Most importantly, the energetic considerations in Sec.~\ref{sec: mechanism}, such as ES energy $\sim |\delta \nu|^{3/2}$, stiffness $\sim|\delta\nu|^{1/2}$ and $|\delta\nu_\text{critical}| \sim \tilde{m}_D^2$ remain qualitatively the same, revealing a potentially general mechanism for exotic (higher-spin) ES.

Here we present another feature that may serve to confirm the higher-spin ES. In the presence of disorder, we recognize that the transitions between different different $C_F$ states are analogous to the QH plateau transitions, such that the transition from the layer decoupled QH state with $C_F=1$ into the ES state with $C_F=-2n$ necessarily take multiple steps such that suitable numbers of extended states can pass the Fermi level of $F$, resulting potential intermediate states before entering the ES phase. The transition boundary into the ES state is between a $C_F =-2n$ and a $C_F=-(2n-1)$ state, thus the counterflow  conductivities at the critical point will take on universal values. Based on the (highly simplified) mean field picture where $F$ are approximated to be non-interacting, we find this critical pseudospin conductivity to be:
\begin{align}
    2\pi \hat{\sigma}^\text{s}_{ij} = 4n^2 \delta_{ij} - (4n^2-2n) \epsilon_{ij}
\end{align}
which could be measured by contour-flow transport measurement. We caution that owing to the strongly coupled nature of the problem, the mean-field approach is of limited reliability~\cite{PhysRevB.109.085143,nosov2025plateau}. However, this approach does capture the fact that the universal conductivities changes systematically with the pseudospin ($2n$) of the exciton condensate. For instance, in the usual superfluid-insulator transition of bosons with charge $q$, the 2D conductivity at criticality $\sigma^{(1)}_{xx}=Aq^2$, where $A$ is a universal constant. Now, if we consider a higher charge condensate, $q'=nq$, then the universal conductivity at that transition will be larger by a factor of $n^2$, i.e. $\sigma^{(n)}_{xx} = n^2\sigma^{(1)}_{xx}$. Generally, once the critical resistivity tensor is known for $F$, which we call $\hat{R}^F$, one can deduce the physical one through
\begin{align}
    \hat{R}^\text{s}= \hat{R}^F +\frac{1}{2n} \begin{bmatrix}
    0 & -1 \\
    1 & 0
\end{bmatrix}
\end{align}

{\bf Comparison with existing experiments: } Our generalized theory for higher Jain fractions can capture some of the  existing experimental observations. In Refs.~\cite{li2019pairing,liu2019interlayer}, the resistivity was measured for $\nu^{(\updownarrow)}=2/5,3/7$ at a relatively small value of $d/\ell_B$ where interlayer correlated states are expected. The measured results for $\nu^{(\updownarrow)}=2/5$ ($n=2$) were:
\begin{align}
    \frac{1}{2\pi}R^\text{drive}_{xy} =& \frac{3}{2} \ , \  \frac{1}{2\pi}R^\text{drag}_{xy} = 1  \ , \  R^\text{drag}_{xx} \approx R^\text{drive}_{xx} \approx 0
\end{align}
which -- within our theory -- implies that $\hat{R}^F_{ij}\approx 0$ for all $ij$, i.e. the $F$ fermions forms a good conducting state. This is consistent with the scenario that $F$ forms a gapless state e.g. by having a Fermi surface. Our theory suggests that further lowering temperature or tuning $d/\ell_B$ from this experimental setup should give rise to a gap for the fermions, leading a different quantization of the resistivity with $\frac1{2\pi} (R^\text{drive}_{xy}-R^\text{drag}_{xx})-\frac12 = \frac2{C_F}$ with a well-defined $C_F$. We further note that Refs.~\cite{li2019pairing,liu2019interlayer} present a compelling alternate scenario for these observations at $\frac25+\frac25$, in terms of an integer QH state of the CF theory we adopted for $\nu_\text{T}=2/3$.

On the other hand, the measured results for $\nu^{(\updownarrow)}=3/7$ ($n=3$) were
\begin{align}
    \frac{1}{2\pi}R^\text{drive}_{xy} =& \frac{5}{3} \ , \ \ \frac{1}{2\pi}R^\text{drag}_{xy} = \frac{2}{3} \ , \  R^\text{drag}_{xx} \approx R^\text{drive}_{xx} \approx 0
\end{align}
which can be readily explained by a $f+\mathrm{i}f$ weak pairing phase (angular momentum $L=3$) between $f^{(\updownarrow)}$ with $C_F=3$. The transition into this phase from the large $d/\ell_B$ phase, which has $p+\mathrm{i}p$ pairing between $f^{(\updownarrow)}$, is thus predicted to necessarily break rotation symmetry to two-fold,  similar to the transition into the $p-\mathrm{i} p$ (with $L=-1$) considered in subsection~\ref{sec: principal}.

\subsection{Other generalizations}
\label{sec: conjugate}

Our theory admits some straightforward generalization to other setups. For example, it can be applied to {\it bosonic} QH systems with little modification. Consider the simplest case of a bilayer of bosons at filling fraction $\nu^{(\updownarrow)}=1/2$. At large layer separation they are expected to form a layer decoupled $(220)$. Then if the lowest energy pseudospin excitation is formed from oppositely charged semions in the two layer, this composite object is a fermion. This is different from the cases considered in this paper where they have carried nontrivial anyonic statistics. Injecting layer imbalance while keeping the total density fixed would then imply doping of fermionic excitons in analogy with~\cite{PhysRevLett.121.026603}. Then we can ask what happens when the two layers are brought into proximity. As we describe below, the same considerations applied to $1/3+1/3$ case predict here a transition into the toric code topological order! To see this one simply needs to change the flux attachment matrix in \ref{eq: LCS} to $\begin{bmatrix}
    1 & 1 \\
    1 & 1
\end{bmatrix}$, 
which still set the CFs $f^{(\sigma)}$ to two composite Fermi seas. Then, the $p\pm \mathrm{i} p$ paring states of them correspond to a $(002)$ and a $(220)$ state, respectively, the former of which is nothing but the toric code topological order. 

Another direction of generalization is the following. Upon a partial charge conjugation for one of the layers, say $\downarrow$ layer, $A^{(\downarrow)}\rightarrow - A^{(\downarrow)}$ and $\psi^{(\downarrow)}\rightarrow \psi^{(\downarrow),\dagger} $. This swaps $A^\text{c}\leftrightarrow A^\text{s}$ and reverses the effective magnetic field on $\downarrow$ layer. Therefore, the problem converts to $\nu^{(\updownarrow)} = 1/3,-1/3$ filling of a pair of Chern bands with {\it opposite} Chern number, which may be realizable in certain moire materials, and the ES states will convert to superconducting states. We note that the resulting wavefunction is similar to that in Ref.~\cite{PhysRevB.111.014508}.

\section{Conclusion and Discussions}

We developed a theory for the continuous transitions between bilayer QH states with different topological orders by considering the band topology transitions of the relevant neutral anyonic excitons. The critical point is described with QED$_3$-CS theory where the anyons admit Dirac dispersion. Upon doping pseudospin carried by the anyons near the criticality, we show that these anyons have strong tendency to form anyon superfluid, signaled by an anomalously soft energy scaling and large phase stiffness. Our theory naturally predicts enhanced  tendencies towards ES and broken translation and rotation symmetries in the vicinity of the transition, which are experimentally testable and may generate new understanding of anyon superfluidity.

QH bilayers with balanced filling $\nu^{(\updownarrow)} = 1/3$, and more generally $\nu^{(\updownarrow)} = \frac{n}{2n+1}$, have been studied in two complementary device geometries.  Li \textit{et al.} fabricated an edgeless Corbino annulus with the graphene layers separated by a $\sim 4.5\,\text{nm}$ hBN tunnel barrier and subjected the system to magnetic fields up to $30\,\text{T}$~\cite{zhang2025excitons,nguyen2025bilayerexcitonslaughlinfractional}.  The absence of edge channels allows bulk transport to be probed directly; under slight layer imbalance the device exhibits nearly perfect Coulomb--drag quantization, consistent with excitonic low--energy degrees of freedom and indicative of exceptional sample quality.  Kim \textit{et al.} employed a conventional Hall bar with a thinner ($\sim 2.5\,\text{nm}$) hBN spacer and operated at fields up to $25\,\text{T}$~\cite{liu2019interlayer}.  In this geometry the counter--flow channel retains an unquantised Hall resistance at many fractional fillings, suggesting dissipative domain walls and highlighting the need for further uniformity improvements.

This contrast underscores a geometry--diagnostic trade--off.  Measurements in the Corbino geometry verifies an incompressible bulk yet cannot reveal the dissipationless neutral supercurrent that defines an exciton superfluid.  A Hall bar is required to demonstrate the ``gold--standard'' condition signaled by pseudospin (so-called counter-flow) resistivity 
\begin{equation}
    R_{xx}^{\text{s}}\to 0,\qquad R_{xy}^{\text{s}}\to 0 ,
\end{equation}
so far achieved only at total filling $\nu_\text{T} = 1/2+1/2$ in graphene double layers~\cite{Liexcitonic2017,Liucrossover2022}.  A slightly weaker but still stringent condition is perfect drag,
\begin{equation}
    R_{xx}^{\text{drag}}\to 0,\qquad \frac{1}{2\pi}R_{xy}^{\text{drag}} = \frac{1}{\nu_\text{T}} ,
\end{equation}
which is likewise satisfied at $\nu_\text{T} = 1/2+1/2$~\cite{Liuquantum2017,zeng2023}.  The latter work extends to large imbalance $\nu^{(\uparrow)} = 1/3$ and $\nu^{(\downarrow)} = 2/3$, where tightly bound bosonic excitons are expected to condense directly~\cite{PhysRevX.13.031023}.  Neither benchmark has yet been met for the balanced $\nu^{(\updownarrow)} = 1/3$ system: the Corbino experiment confirms only a bulk gap~\cite{zhang2025excitons,nguyen2025bilayerexcitonslaughlinfractional}, and existing Hall bars still display drag dissipation~\cite{liu2019interlayer}.  Progress therefore hinges on fabricating next--generation high--mobility Hall bars capable of demonstrating $R_{xx}^{\text{s}} = R_{xy}^{\text{s}} = 0$, thereby establishing the anyonic exciton superfluid proposed here.

We now return to the balanced bilayer and focus on the $(330)\!\to\!(112)$ transition, and more generally on its counterparts in the higher Jain fractions. In one of the experiments by Li \textit{et al.}~\cite{nguyen2025bilayerexcitonslaughlinfractional}, the two phases are separated by a single jump in transport as the magnetic field—and hence $d/\ell_{B}$—is swept. In principle only the neutral counter-flow sector should become critical, with the charge gap remaining intact.  Curiously, the Hall-bar data in this experiment also show enhanced parallel‐flow conduction at the putative critical field; this incongruity may be due to unexpected mixing between parallel-flow and counter-flow signals, and calls for further systematic study. Besides the characteristic transport signals, we note that a continuous transition between these two states is also accompanied by a power-law layer polarizability at the critical point; that is, the layer density imbalance $P$ should scale with a power of the applied perpendicular displacement field $D_z$; within our theory this should be $P\sim D_z^2$.

Because the $(330)$ and $(112)$ states possess different shifts and, in our CF picture, differ by a $p_{x}\!\pm\!\mathrm{i} p_{y}$ pairing channel, the critical point must break the native rotational symmetry down to $C_{2}$.  Li \textit{et al.} have recently developed an angle-resolved transport technique that can directly test this nematic prediction.  By placing multiple voltage probes at distinct in-plane orientations on an edgeless Corbino sample, they extract the full symmetric‐flow resistivity tensor $\hat{R}_{ij}(\theta)$ in different directions~\cite{LiAngle1,LiAngle2,LiAngle3}.  Observation of a pronounced two-fold anisotropy emerging precisely at the $(330)\!\leftrightarrow\!(112)$ critical field would provide a decisive confirmation of the composite-fermion critical theory advanced here. Another notable prediction is the observation of an enhanced critical temperature of the ES state on pseudospin doping in the vicinity of  such a critical point.

Our theory provide several concrete predictions to higher Jain fractions $\nu^{(\updownarrow)}=n/(2n+1)$, which can be readily examined by the above experimental protocols. Most prominently we predict direction topological phase transitions could arise as tuning $d/\ell_B$ similar to the cases in $\nu^{(\updownarrow)}=1/3$, which will generically be accompanied by a rotation symmetry breaking. Upon doping pseudospin near the criticality, our theory suggests the general appearance of pseudospin-$2n$ ES through an anyon mechanism, thus providing a novel proposal for the realization of spin nematic phase. We note that these higher-pseudospin condensation can be in principle confirmed through signatures in Josephson tunneling and Andreev reflection, in means analogous to the procedures of identifying charge-$4e$ superconductors~\cite{berg2009charge,PhysRevB.109.144504}.

{\bf Acknowledgement.} We thank Bertrand I. Halperin, Steven A. Kivelson, J.I.A. Li, Naiyuan J. Zhang, Pietro Bonetti, Da-Chuan Lu, Xue-Yang Song, Pavel A. Nosov, Carolyn Zhang, Seth Musser and Darius Zhengyan Shi for helpful discussions. This work is supported by the Simons Investigator award, the
Simons Collaboration on Ultra-Quantum Matter, which
is a grant from the Simons Foundation (651440, A.~V.; 1151944, M.~Z.).

\bibliography{ref}

\appendix
\onecolumngrid

\section{Parton construction}
\label{app: parton}

Here, we use the parton approach to understand the flux attachment scheme of the specific composite fermion (CF) theory used in the main text, and how it generalizes to other filling fractions of interest. We adopt parton decomposition:
\begin{align}
    \psi^{(\sigma)} = c \cdot  c^{(\sigma)} \cdot f^{(\sigma)}
\end{align}
where $c,c^{(\sigma)} $ are auxiliary femionic parton fields, and $f^{(\sigma)}$ are the CF fields. We assume that the auxiliary partons are well-gapped such that they can be integrated out, and we seek ansatz states for these partons such that the CF fields see no net flux on average at the relevant `undoped' fillings.

Three dynamical \spinc connections, $a^{(\updownarrow)}$ and $b$, are needed for gluing up $c$, $c^{(\sigma)}$ and $f^{(\sigma)}$ in an effective field theory description. (They are \spinc because their sources are fermionic fields.) Formally, we can then write
\begin{align}
    \mathcal{L}_\psi = \sum_{\sigma}\left\{ \mathcal{L}_f [f^{(\sigma)};a^{(\sigma)}] + \mathcal{L}'[c^{(\sigma)}; -a^{(\sigma)} + A^{(\sigma)}-b]\right\} +\mathcal{L}''[c;b]
\end{align}
where ``$;$'' symbols separate the parton degrees of freedom and the gauge fields they couple to. 

Based on this decomposition, various fermionic parton (FP) theories valid at different filling fractions can be obtained in a unified framework, depending on what QH states we put the auxiliary partons in. Specifically, if we put $c^{(\updownarrow)}$ to QH states at effective filling fractions $\nu'$ and $c$ to QH state at effective filling fraction $\nu''$, the balanced electron filling fractions at which such FP theories are relevant (in the sense that the FP fields see no net flux) can be obtained as
\begin{align}
\nu^{(\updownarrow)} = \frac{\nu' \nu''}{2\nu '+ \nu'' }
\end{align}
Below, we discuss several physically relevant FP theories.

\subsection{$\nu'$ integer, $\nu''=1$: the principal Jain sequence}

If we put $c^{(\sigma)}$, $c$ in IQH states, at filling fractions $\nu'^{(\sigma)}=n$ and $\nu''=1$ respectively, with $n$ arbitrary non-zero integers, the electron filling fractions are 
\begin{align}
\nu^{(\updownarrow)} = \frac{n}{2n+1}
\end{align}
which includes the fractions in the principal Jain sequence. The effective field theory becomes
\begin{align}
     \mathcal{L}_\psi &= \sum_{\sigma}\mathcal{L}_f [f^{(\sigma)};a^{(\sigma)}]  + \mathcal{L}_\text{CS}\\
     \mathcal{L}_\text{CS} & =   \sum_{\sigma,I=1..n} \left\{-\frac{1}{4\pi} \alpha^{(\sigma,I)}\mathrm{d}\alpha^{(\sigma,I)} + \frac{1}{2\pi} \left[A^{(\sigma)}-a^{(\sigma)}-b + \left(I-\frac{1}{2}\right) \omega \right] \mathrm{d}\alpha^{(\sigma,I)}\right\}-\frac{1}{4\pi} \beta\mathrm{d}\beta + \frac{1}{2\pi} \left(b + \frac{1}{2}\omega \right)\mathrm{d}\beta \label{eq: L CS bare}
\end{align}
where $\alpha^{(\updownarrow,I)}$ and $\beta$ are ordinary $U(1)$ gauge fields. Integrating out $b$, we find it simply enforces $\beta =\sum_{\sigma,I=1...n} \alpha^{(\sigma,I)}$, and then the Lagarangian simplifies to
\begin{align}
     \mathcal{L}_\text{CS}  =&  \sum_{\sigma,I=1,\dots,n}  \frac{1}{2\pi} \left(A^{(\sigma)}-a^{(\sigma)} + I \omega \right) \mathrm{d}\alpha^{(\sigma,I)}- \frac{1}{4\pi}
\vec{\alpha}^{T}  ({\bf I }+  {\bf E })_{2n\times 2n}  \mathrm{d}
  \vec{\alpha}\nonumber\\
 \vec{\alpha} \equiv & \begin{pmatrix}
\alpha^{(\uparrow,1)} & \dots & \alpha^{(\uparrow,n)}  & \alpha^{(\downarrow,1)} & \dots & \alpha^{(\downarrow,n)}
\end{pmatrix}^T
\end{align}
where ${\bf I}$ is the identity matrix and ${\bf E}$ represents a matrix with all entries equal to $1$. In particular, choosing $n=1$ recovers the CF theory in Eq.~\ref{eq: LCS} that applies to the case of $\nu^{\updownarrow}=1/3$ studied in the main text.

The equations of motion of $a^{(\sigma)}$ and $\alpha^{(\sigma,I)}$ imply flux attachment scheme:
\begin{align}
    \frac{\nabla\times\bm{\alpha}^{(\sigma,I)}}{2\pi} & = \rho^{(\sigma)}/n \\
    \frac{\nabla\times\bm{a}^{\sigma}}{2\pi} & = \frac{\nabla\times\bm{A}^{(\sigma)}}{2\pi} - \rho^{(\uparrow)}-\rho^{(\downarrow)}-\rho^{(\sigma)}/n
\end{align}
It can be seen that for $|n|\neq 1$, the FP $f^{(\sigma)}$ are generically not CFs since they have fractional flux attachment. 

An alternative way of rewriting $\mathcal{L}_\psi$ can be obtained by integrating out $\alpha$ and $\beta$ fields of Eq.~\ref{eq: L CS bare}. This amounts to
\begin{align} \label{app: eq: LCS alter}
    \mathcal{L}_\text{CS} 
    & = \sum_{\sigma}\frac{n}{4\pi} \left(a^{(\sigma)}-A^{(\sigma)}+b\right)\mathrm{d} \left(a^{(\sigma)}-A^{(\sigma)}+b\right)   + \frac{1}{4\pi} b\mathrm{d}b  -\sum_{\sigma} \frac{n^2}{2} \frac{1}{2\pi} \omega \mathrm{d} a^{(\sigma)}+\frac{n^2}{2\pi} \omega \mathrm{d} A^\text{c}\nonumber\\
    &\ \ \ \ \ -\left(n^2 -\frac{1}{2}\right)\frac{1}{2\pi} \omega\mathrm{d}b  +\frac{8n^3-2n+3}{12} \frac{1}{4\pi} \omega \mathrm{d} \omega   + \left(2n+1\right)\Omega_g
\end{align}
This expression is useful when considering the effective theory of paired states between $f^{(\uparrow)}$ and $f^{(\downarrow)}$. Specifically, consider the hydrodynamical theory of a pair field with angular momentum $L$,
\begin{align}\label{app: eq: L Delta eff}
\mathcal{L}_{\Delta, \text{eff}} = \frac{1}{2\pi}  \left(a^{(\uparrow)}+a^{(\downarrow)}+L \omega\right) \mathrm{d} \eta .
\end{align}
then integrate out $\eta$ and take the parametrization
\begin{align}
    a^{(\uparrow)} + \frac{L}{2} \omega  = -  a^{(\downarrow)} -\frac{L}{2}\omega \equiv a.
\end{align}
We can then derive an alternative expression for the combination 
\begin{align}  \label{app: eq: L CS'}
    \mathcal{L}_\text{CS}' 
 =& \frac{2n }{4\pi}  a\mathrm{d} a  -\frac{2n}{2\pi}A^\text{s} \mathrm{d} a + \frac{2n}{4\pi}  A^\text{s}\mathrm{d} A^\text{s}   \nonumber\\
    & 
    +   \frac{2n+1}{4\pi}  b\mathrm{d} b -\frac{1}{2\pi} \left[2n A^\text{c}+\left(n^2-\frac{1}{2}+nL\right)\omega\right] \mathrm{d}b  + \frac{2n}{4\pi}  A^\text{c}\mathrm{d} A^\text{c} \nonumber\\
    &   +\frac{n(n+L)}{2\pi}A^\text{c}\mathrm{d}\omega + \left(\frac{8n^3-2n+3}{12}+\frac{nL^2}{2} +n^2L\right)\frac{1}{4\pi}\omega\mathrm{d}\omega+\left(2n+1 \right)\Omega_g
\end{align}

We note that, when the rotation symmetry is broken down to $C_m$, then only the mod-$m$ part of the angular momentum $L$ is well defined. That is because now a $\frac{m}{2\pi}\omega\mathrm{d}\eta^\text{r}$ term needs to be added, but the rotational hydrodynamic field $\eta^\text{r}$ can be arbitrarily redefined by absorbing $\eta$ in Eq.~\ref{app: eq: L Delta eff} which changes the coefficient of the $\frac{1}{2\pi} \omega \mathrm{d} \eta$ by $m$. This means that only the mod-$m$ part of $L$ is well defined.

\subsection{$\nu' = 1/(2p-1), \nu''=1/(2q+1)$: Laughlin sequence}

If we put $c^{(\sigma)}$, $c$ in the Laughlin states, at filling fractions $\nu'^{(\sigma)}= 1/(2p-1)$ and $\nu'' = 1/(2q+1)$ with $p,q$ arbitrary integers, the corresponding electron filling fractions are 
\begin{align}
\nu^{(\updownarrow)} = \frac{1}{4q+2p+1}
\end{align}
which includes all the fractions in the Laughlin sequence. The effective field theory becomes:
\begin{align}
     \mathcal{L}_\psi &= \sum_{\sigma}\mathcal{L}_f [f^{(\sigma)};a^{(\sigma)}]  + \mathcal{L}_\text{CS}\\
     \mathcal{L}_\text{CS} & =   \sum_{\sigma} \left[-\frac{2p-1}{4\pi} \alpha^{(\sigma)}\mathrm{d}\alpha^{(\sigma)} + \frac{1}{2\pi} \left(A^{(\sigma)}-a^{(\sigma)}-b + \frac{2p-1}{2} \omega \right) \mathrm{d}\alpha^{(\sigma)}\right]-\frac{2q+1}{4\pi} \beta\mathrm{d}\beta + \frac{1}{2\pi} \left(b +\frac{2q+1}{2}\omega \right)\mathrm{d}\beta
\end{align}
where $\alpha^{(\updownarrow)}$ and $\beta$ are ordinary $U(1)$ gauge fields. Integrating out $b$, we find it simply enforces $\beta =\alpha^{(\uparrow)}+\alpha^{(\downarrow)}$, and then the Lagarangian simplifies to
\begin{align}
     \mathcal{L}_\text{CS} & =  \frac{1}{2\pi} \left[A^{(\sigma)}-a^{(\sigma)} + (p+q)\omega \right] \mathrm{d}\alpha^{(\sigma)}- \frac{1}{4\pi}\begin{pmatrix}
\alpha^{(\uparrow)} & \alpha^{(\downarrow)}
\end{pmatrix} \begin{bmatrix}
2p+2q & 2q+1 \\
2q+1 & 2p+2q
\end{bmatrix} \begin{pmatrix}
\mathrm{d}\alpha^{(\uparrow)} \\ \mathrm{d}\alpha^{(\downarrow)}
\end{pmatrix}
\end{align}
We thus see that the filling fractions correspond to Halperin $[(2p+2q+1)(2p+2q+1)(2q)]$ or $[(2p+2q-1)(2p+2q-1)(2q+2)]$ state, thus the $p\pm \mathrm{i}p$ pairing transition of the CFs may directly describe their transitions. 

In particular, choosing $p=1$, $q=0$ recovers the case of $\nu^{(\updownarrow)}=1/3$ studied in the main text.

\section{Two shortcuts to the topological properties of the exciton superfluid}

\subsection{$K$-matrix approach}

\label{app: smith}

Here we use the $K$-matrix formalism to infer the properties of the anyonic ES when doping the $(330)$ state, which is described by the topological theory:
\begin{align}\label{eq: LCS'}
    \mathcal{L}_{(330)} &= \sum_\sigma \frac{1}{2\pi}  A^{(\sigma)} \mathrm{d}\alpha^{(\sigma)}  - \frac{1}{4\pi}\begin{pmatrix}
\alpha^{(\uparrow)} & \alpha^{(\downarrow)}
\end{pmatrix} 
\begin{bmatrix}3 & 0 \\
0 & 3
\end{bmatrix} \begin{pmatrix}
\mathrm{d}\alpha^{(\uparrow)} \\ \mathrm{d}\alpha^{(\downarrow)}
\end{pmatrix}
\end{align}

We describe the relevant charge-neutral but pseudospin-carrying anyons with a bosonic field $\phi$ minimally coupling to $\alpha^{(\uparrow)}-\alpha^{(\downarrow)}$:
\begin{align}
    \mathcal{L} = \mathcal{L}_{(330)}+ \mathcal{L}_\phi[\alpha^{(\uparrow)}-\alpha^{(\downarrow)}]
\end{align}
In order to describe pseudospin doping it is convenient to consider introducing fermions. We therefore perform a further fermion-boson duality, with a fermionic field $\psi$ ($c$ is \spinc)
\begin{align}
    \mathcal{L} = \mathcal{L}_{(330)}+ \mathcal{L}_\psi[c+\alpha^{(\uparrow)}-\alpha^{(\downarrow)}] +\frac{1}{4\pi} c\mathrm{d} c + \Omega_g
\end{align}

From the equations of motion of $\alpha^{(\updownarrow)}$, we have
\begin{align}
\frac{1}{2\pi}\langle \nabla \times \bm{\alpha}^{(\sigma)}\rangle =& \frac{\sigma }{3}\rho_\psi \\
\frac{1}{2\pi}\langle \nabla \times  \bm{c}\rangle =& -\rho_\psi
\end{align}
which means that the $\phi$ particles are at an effective filling fraction $\nu_\psi = -3$. Putting the $\psi$ particles into an integer QH state, we reach
\begin{align}
    \mathcal{L} = & \mathcal{L}_{(330)}  + \frac{1}{4\pi} c \mathrm{d} c  + \Omega_g - \frac{3}{4\pi}  \left(\alpha^{(\uparrow)}-\alpha^{(\downarrow)} +c\right)\mathrm{d} \left(\alpha^{(\uparrow)}-\alpha^{(\downarrow)} +c\right)   - 3\Omega_g \\
 = & \sum_\sigma \frac{1}{2\pi} A^{(\sigma)} \mathrm{d}\alpha^{(\sigma)}  - \frac{1}{4\pi}\begin{pmatrix}
\alpha^{(\uparrow)} & \alpha^{(\downarrow)} & c
\end{pmatrix} \begin{bmatrix}
6 & -3 & 3\\
-3 & 6 & -3\\
3 & -3 & 2
\end{bmatrix} \mathrm{d}\begin{pmatrix}
\alpha^{(\uparrow)} \\ \alpha^{(\downarrow)} \\ c
\end{pmatrix}  -2\Omega_g
\end{align}

The K-matrix above has vanishing determinant, hence it is useful to implement a unimodular transformation 
\begin{align}
    \begin{pmatrix}
\tilde{a} \\ \tilde{b} \\ \gamma
\end{pmatrix}  = 
    \begin{bmatrix}
1 & -2 & 1\\
-2 & 1 & -1\\
-1 & 0 & 0
\end{bmatrix} \begin{pmatrix}
\alpha^{(\uparrow)} \\ \alpha^{(\downarrow)} \\ c
\end{pmatrix} 
\end{align}
the theory becomes (here $\tilde{a},\tilde{b}$ are \spinc, and $\gamma$ is an ordinary $U(1)$ gauge field):
\begin{align}
    \mathcal{L} =  & - \frac{2}{2\pi} A^{\text{s}}\mathrm{d}\gamma -2\Omega_g \nonumber\\
    &+\frac{1}{2\pi} \left(-A^{\text{c}}+A^{\text{s}} \right) \mathrm{d}\left(\tilde{a}+\tilde{b}\right) \nonumber\\
 & - \frac{1}{4\pi}\begin{pmatrix}
\tilde{a} & \tilde{b}
\end{pmatrix} \begin{bmatrix}
2 & 1 \\
1 & 2 \\
\end{bmatrix} \mathrm{d} \begin{pmatrix}
\tilde{a} \\ \tilde{b}
\end{pmatrix} 
\end{align}
So we find that this theory describes a pseudospin-$2$ ES with a residual topological order having topological degeneracy $g=3$ and charge Hall coefficient $2\pi \sigma^\text{c}_\text{H}= 2/3$. The chiral central charge is $c_-=0$.

\subsection{Stack-and-condense approach}
\label{app:sec: stack&condenseHierarchy}
Here, we outline a shortcut for getting the properties of the anyon exciton condensate, such as its residual topological order, the condensate charge and edge modes. The route to obtaining these properties will be through the mechanism of stacking followed by anyon condensation, similar to what was used for the quantum Hall hierarchy and daughter states in a recent work~\cite{zhang2025hierarchyconstructionnonabelianfractional}. 

As an example, we now analyze the anyonic ES obtained on doping the $(1/3, -1/3)$ charged anyons, denoted $\bf{a}$, into the $(330)$ state which we denote as $\mathcal{C}$.  Note, the self-statistics angle of $\bf{a}$ is $\theta_{\bf{a}} = 2\pi/3$, and the chiral central charge to begin with is $c_-=2$. Now, we consider the effect of doping excitons to be modeled by first attaching a charge and pseudospin {\it neutral} (i.e. not coupling to $A^{(\sigma)}$) topological order ${\mathcal D}$. In principle, one can always allow for such a state to be supplied, and when properly selected, can mimic the result of a finite density of nontrivial anyons. For our purposes, we select ${\mathcal D}$ to be a $(\bar{2}\bar{2}\bar{1})$ bosonic Laughlin state with $\mathbb{Z}_3$ type topological order described by a K matrix $K' =- \begin{bmatrix} 
    2  & 1\\
    1   & 2
    \end{bmatrix}$.
[Note, the $K=3$ topological order does not satisfy the spin-charge relation, so that it cannot emerge in condensed matter systems.] For this stacked topological order, the chiral central charge $c_-=2$ and the anyons can be labelled by the composition vector $\vec{l}= ( l_1,\, l_2) $. The mutual statistics bewteen two anyons with compositions $\vec{l}$, $\vec{l}'$ is given by $\theta_{\vec{l},\vec{l}'}  = \pi \vec{l}\cdot K'^{-1}\cdot \vec{l} = -\frac{\pi}{6} [(l_1+l_2)(l_1'+l_2')+3(l_1-l_2)(l_1'-l_2')]$. In particular the anyon with $\vec{l}=(1,1)$ has self-statistical phase $\theta_{\vec{l}}=-2\pi/3$.

We now consider the stacked theory $\mathcal{C} \boxtimes \mathcal{D}$. Now, the resulting state has total central charge $c_-=0$ and we can find a composite object that is a bound state of $\vec{l}=(1,1)$ and the anyon $\bf{a}$, which is now a self-boson. We consider condensing this boson. the edge central charge does not change in this process and remains $c_-=0$, while it is easily seen that all the anyons in $\mathcal{D}$ as well as all the charge-neutral, pseudospin-carrying anyons in $\mathcal{C}$ are confined in this process, leaving behind a topological order purely in the charge sector. Finally, since we condensed a anyon with pseudospin-$2/3$, we have an exciton condensate. However, since this fractional pseudospin boson is not a local excitation, but its third power is, we have a psuedospin $2$ condensate coexisting with the $(112)$ topological order. Finally, we can deduce that the shift of this state is the same as the $(330)$ state. This follows because the shift can be viewed as a mixed symmetry-enriched response between rotation symmetry and charge. Neither of these symmetries are touched in the stack and condense process, where all relevant low energy anyons are charge neutral. Thus the shift is left invariant and is inherited from the parent state.

\section{The role of the symmetries}
\label{sec: symmetry}
In this section we discuss the roles of the symmetries in this system. 

\subsection{Rotation symmetry} 

The $SO(2)$ or a $C_{2m}$ ($m>1$) rotation symmetry ensures that the $p_x \pm \mathrm{i}p_y$ pairing modes belong to different irreducible representations, and thus cannot mix. That ensures the form of the effective Lagrangian $\mathcal{L}_\Delta$ in Eq.~\ref{eq: L Delta}. Any coexistence of the two pairing modes, which is necessary for a direct transition between these two, must break the symmetry to $C_2$ at mean field level, which has been understood by the existence of a soft direction in the BdG spectrum. To see this from the $\Delta$ sector, it is convenient to parameterize 
\begin{align}\label{eq: Delta}
    \vec{\Delta} = |\vec{\Delta}| \mathrm{e}^{\mathrm{i}\varphi}\left(\cos\frac{\vartheta}{2} \mathrm{e}^{\mathrm{i}\varrho} , \sin\frac{\vartheta}{2} \mathrm{e}^{-\mathrm{i}\varrho}\right)
\end{align}
The $\varphi$ is the usual superconducting phase that couples to the gauge fields, $\vartheta$ determines the composition of the pairing fields, and $\varrho$ is a directional angle that transforms to $\varrho +\gamma$ under a spatial rotation by $\gamma$ and thus couples to the spin connection. As long as $\sin \vartheta \neq 0$, the rotation symmetry of the system breaks down to $\mathcal{I}$ by the saddle-point solution. The residual $\pi$ rotation ($C_2$) symmetry is due to the fact that the phase variables $(\varphi,\varrho)$ are periodic under shift $(\pi,\pi)$ so $\varrho$ and $\varrho +\pi$ should be regarded as the same up to gauge transformation in $\varphi$.

What has been neglected in the previous discussion is the fluctuation of the nematic phase $\varrho$. In the case of spontaneouly broken $SO(2)$/$C_{2m}$ rotation symmetry, this fluctuation leads to a $\bm{k}^2$ gapless/gapped phason mode at long distance.  However, there is another possibility when its fluctuation is too wild such that the rotation symmetry is restored, causing only the product $\Delta^{(+)}\Delta^{(-)}$ or its higher powers to condense. In the context of superconductivity, such orders are called charge-$2ke$ ($k>1$) superconducitivity and can be formed as vestigial orders of multi-component superconductors~\cite{berg2009charge,jian2021charge,PhysRevLett.127.047001}. Such states should have enlarged topological order, the nature of which will be left for future study.

For a continuous transition between $p_x \pm \mathrm{i}p_y$, it is crucial to have the remnant $C_2$ symmetry. Otherwise, the direct transition line represented by $3$ in Fig.~\ref{fig: CF phase diagram} should splits into two and sandwich an intermediate phase with trivial band topology that connects to the strong pairing phase.

\subsection{Inter-layer exchange} 

It is possible that the two layers may be distinct and thus have different CF dispersions. However, as long as the $C_2$ symmetry is present, the direct transitions discussed in this work are still possible. What might be interesting in this scenario is that the pairing mode now may have non-zero center-of-mass momentum through a FFLO mechanism~\cite{annurev:/content/journals/10.1146/annurev-conmatphys-031119-050711}, forming exotic versions of bilayer QH states. The absence of interlayer exchange symmetry also removes the symmetry between $\delta \nu \leftrightarrow -\delta \nu$ upon doping.

\subsection{Pseudospin conservation}

When the two layers are not well insulated, interlayer tunnelings are possible which breaks the pseudospin conservation. In this case, the different pairing modes with different pseudospin can mix. In the current context, the interlayer pairing modes which has pseudospin $0$ should generically mix with the intralayer conterparts with psesdospin $\pm 2$. These will split each the two BdG bands of complex fermions into four BdG bands of majorana fermions, and again split the transition between $p_x\pm \mathrm{i} p_y$ weak pairing phases into two transitions, but more crucially also split the transition from weak $p_x\pm \mathrm{i} p_y$ to strong pairing, where there could be an intermediate phase where the occupied majorana bands have an odd Chern number. The corresponding QH states include an inter-layer Pfaffian state~\cite{PhysRevB.61.10267,PhysRevB.91.205139,PhysRevB.65.041305}, and such states hosts a non-Abelian Ising anyons similar to the Moore-Read state~\cite{moore1991nonabelions}. Interestingly, in the numerical study in Ref.~\cite{PhysRevB.91.205139}, a phase diagram consisting of $(112)/(330)$, interlayer Pfaffian, and cooper pair bosonic QH phases in a sequential order has been found upon including explicit interlayer  attractions, which might be consistent the picture of an emergent effective inter-layer tunneling. We also note that other proposals for the phase diagram upon including the inter-layer tunneling include various other possible phases~\cite{PhysRevLett.113.236804}.

\section{Estimation of superfluid stiffness}
\label{app: sec: stiffness}
Here we estimate the superfluid stiffness for the ES obtained via doping the critical point. For concreteness let's focus on the filling fraction $1/3+1/3$ and assume there is only one Dirac node near the criticality. The generalizations are straightforward. 

First let's briefly review what this task amounts to within the FP theory. We start from the simplified action:
\begin{align}
\mathcal{L}=\mathcal{L}_F[a]+ \frac{2 }{4\pi}  a\mathrm{d} a  -\frac{2}{2\pi}A^\text{s} \mathrm{d} a + \dots
\end{align}
After integrating out $F$ with putting them in a suitable topological state, the effective Lagrangian density becomes
\begin{align}\label{app: eq: gradient expansion}
\mathcal{L}=\frac{1}{2}a_\mu \Pi^{\mu\nu} a_\nu+ \frac{2 }{4\pi}  a\mathrm{d} a  -\frac{2}{2\pi}A^\text{s} \mathrm{d} a + \dots
\end{align}
where  
\begin{align}
    \Pi^{\mu\nu}(x,x') = \langle J^{\mu}(x)J^{\nu}(x')\rangle_F
\end{align}
is a the current-current correlator for the $F$ particles. Then, performing gradient expansion of $\mathcal{L}$ to the second order of $a$, we find the action can generically be parametrized into the form:
\begin{align}\label{app: eq: parametrization}
    \mathcal{L}=\frac{1}{2(2\pi)^2\kappa}(\partial_x a_y -\partial_y a_x)^2 + \frac{1}{2(2\pi)^2\kappa_x}(\partial_t a_y -\partial_y a_t)^2 +\frac{1}{2(2\pi)^2\kappa_y} (\partial_t a_x -\partial_x a_t)^2  -\frac{2}{2\pi}A^\text{s} \mathrm{d} a + \dots
\end{align}
which respects the gauge invariance of $a$. (Here $x,y$ are assumed to be the eigen directions of the anisotropy.) Note that the CS term is canceled assuming the $F$ particles form a IQH state according to the effective filling fraction. 

Since $a$ is a compact $U(1)$ gauge field, it admits an exact re-writing with $J^\mu = \epsilon^{\mu\nu\rho}\frac{1}{2\pi} \partial_\nu a_\rho$ subject to $\partial_\mu J^\mu=0$. With this rewriting we find the action becomes:
\begin{align}
    \mathcal{L}=\frac{1}{2\kappa}J_0^2 - \frac{1}{2\kappa_x}J_x^2 - \frac{1}{2\kappa_y} J_y^2 - 2A^\text{s}_\mu J^\mu + \partial_\mu \theta J^\mu   + \dots
\end{align}
where $\theta$ is a $U(1)$ Lagrangian multiplier enforcing the divergence-free constraint for $J$. Integrating out $J$, we find that the action becomes:
\begin{align}
    \mathcal{L}=\frac{\kappa}{2}(\partial_t \theta-2A^\text{s}_t)^2 - \frac{\kappa_x}{2}(\partial_x \theta-2A^\text{s}_x)^2 - \frac{\kappa_y}{2} (\partial_y \theta-2A^\text{s}_y)^2  + \dots
\end{align}
which agrees with the effective field theory of a ``charge-$2$ superconductor''. We thus recognize that $\kappa$ is the compressibility and $\kappa_{x,y}$ are the superfluid stiffness in the two eigen directions.

Comparing Eq.~\ref{app: eq: gradient expansion} and Eq.~\ref{app: eq: parametrization}, we recognize that
\begin{align}
    \frac{1}{\kappa_x} =&\lim_{q_y\rightarrow 0}\frac{(2\pi)^2}{q_y^2}\Pi^{tt}(q_x=0,q_y,\omega=0)  \\
    \frac{1}{\kappa_y} =&\lim_{q_x\rightarrow 0}\frac{(2\pi)^2}{q_x^2}\Pi^{tt}(q_x,q_y=0,\omega=0)
\end{align}
Therefore, from the long wavelength behavior of the density-density correlation of $F$, we can extract the superfluid stiffness of the resulting ES.

With this understanding, we now evaluate the density density correlation within the mean field theory of FP $F$ near a topological transition. In the main text, we have seen that the low-energy effective theory for $F$ is given by a Dirac theory $\mathcal{L}_\text{Dirac}$. For convenience, let's write down the Hamiltonian (in the basis $\Psi \equiv \mathrm{e}^{\mathrm{i} 2\pi /(3\sqrt{3})(\sigma^x+\sigma^y+\sigma^z)}F$ for clarity):
\begin{align}
    \mathcal{H } = \Psi^\dagger \left[\xi v_x \mathrm{i}D_x \sigma^x -v_y\mathrm{i}D_y \sigma^y  + m_\text{D} \sigma^z\right] \Psi \sim \Psi^\dagger \begin{bmatrix}
        m_\text{D} & \sqrt{2}\bar{v} \hat{a}/\ell\\
        \sqrt{2}\bar{v} \hat{a}^\dagger /\ell & -m_\text{D}
    \end{bmatrix} \Psi
\end{align}
where $\ell\equiv 1/\sqrt{\langle\nabla\times \bm{a}\rangle}$ is the effective magnetic length, $\bar{v} \equiv\sqrt{v_xv_y}$ is the averaged Fermi velocity of the Dirac cone, and 
\begin{align}
    \hat{a} \equiv \frac{\ell}{\sqrt{2}}\left[\sqrt{\frac{v_x}{v_y}} \mathrm{i} D_x + \sqrt{\frac{v_y}{v_x}}D_y\right]
\end{align}
is a ladder operator satisfying $[\hat{a},\hat{a}^\dagger] = 1$. The spectrum and eigenstates are well known for this system. The $n$-th LL is located at ($\text{sgn}(0) \equiv \xi$):
\begin{align}
    \epsilon_n = \text{sgn}(n)\sqrt{m_\text{D}^2 + 2n \bar{v}^2/\ell^2} 
\end{align}
and the single-particle eigenstates therein have the form (taking Landau gauge, $\bm{a}=(0,Bx)$)
\begin{align}
    |n,p_y\rangle \sim  \mathrm{e}^{\mathrm{i}p_y y} \begin{pmatrix}
       \cos\theta_n \cdot \phi_{|n|-1}(\tilde{x}) \\\sin\theta_n \cdot \phi_{|n|}(\tilde{x})
    \end{pmatrix}
\end{align}
where
\begin{align}
    \tilde{x} \equiv& \sqrt{\frac{v_y}{v_x}} \left(x+p_y \ell^2\right) \\
    \tan \theta_n \equiv & \frac{\text{sgn}(n)\sqrt{2n} \bar{v}/\ell}{E_n + m_\text{D}}
\end{align}
and $\phi_n(x)$ is the $n$-th eigenstate of a quantum harmonic oscillator with spread $\ell$. With these understanding, we can compute the (connected piece of) density density correlator at arbitrary frequency and momentum:
\begin{align}\label{eq: Lehmann}
    \Pi^{tt}(\bm{q},\omega)= \int_{p_y,p_y'}\sum_{n\leq n_0, m>n_0}\frac{\langle n,p_y|\mathrm{e}^{\mathrm{i}\bm{q}\cdot \bm{r} }|m,p_y'\rangle\langle m,p_y' | \mathrm{e}^{-\mathrm{i}\bm{q}\cdot \bm{r} }|n,p_y\rangle}{\epsilon_n-\epsilon_m-\omega}
\end{align}
where $n_0$ is the highest occupied LL's index. { When there are multiple Dirac cones, we note that their contribution should be added, but there is no cross term in the $\bm{q}\rightarrow 0 $ limit.  It is a bit tedious but not complicated to evaluate the above expression. } For transparency, we will not present the full results here. A key observation is that for a virtual process involving $n,m$ LLs (assume $|m|>|n|$), the contribution at long wavelength and zero frequency takes the form:
\begin{align}
    \frac{1}{\epsilon_n-\epsilon_m}\frac{1}{2\pi\ell^2}&\left[\frac{\ell^2 }{2}\left(\frac{v_x q^2_x}{v_y}+\frac{v_y q^2_y}{v_x}\right)\right]^{|m|-|n|}  \left[\sin\theta_n\sin\theta_m\sqrt{\frac{|n|!}{|m|!}}L^{|m|-|n|}_{|n|}\left(0\right)  + \cos \theta_n \cos \theta_m\sqrt{\frac{(|n|-1)!}{(|m|-1)!}}L^{|m|-|n|}_{|n|-1}\left(0\right)\right]^2
\end{align}
where $L$ is the generalized Laguerre polynomial. So in order to extract $\kappa_{x,y}$, it suffices to keep the terms involving two LLs with $|m|-|n|=1$. This gives a simple estimation of the superfluid stiffness in the light doping $|\delta\nu|\rightarrow 0$ limit:
\begin{align}
\kappa_{i} \sim \frac{v_\parallel}{v_\perp } (\epsilon_{n_0+1} - \epsilon_{n_0}) \sim \begin{cases}
     \frac{v_\parallel}{v_\perp } \frac{\bar{v} }{\ell } \sim |\delta\nu|^{1/2} \frac{v_\parallel}{v_\perp } \bar{v} \sqrt{\bar{B}}   & m_\text{D} =0 \\
    \frac{v_\parallel}{v_\perp } \frac{\bar{v}^2 }{2m_\text{D}\ell^2 }  \sim |\delta\nu|\frac{v_\parallel}{v_\perp } \frac{\bar{v}^2\bar{B} }{2m_\text{D} }   & m_\text{D} \neq 0
\end{cases}
\end{align}
where $\perp, \parallel$ represent the perpendicular and parallel direction to $i=x,y$. 

{ 
In the main text we have discussed a subtlety of the case of doping $(330)-(112)$ transition point: there is a spontaneous splitting of the LLs at Fermi level. As we have seen above in the Lehmann representation Eq.~\ref{eq: Lehmann}, the stiffness is roughly determined by the direct gap of $F$, and at criticality it is proportional to $|\delta\nu|^{1/2}$. Therefore, the weak splitting of LLs, which only modify the direct gap by an amount $\sim |\delta\nu|$, does not alter the primary scaling of the stiffness. 
}

It is interesting to compare the scaling behaviors of stiffness and that of the compressibility. In the main text we have evaluated that $E(\delta \nu) \sim |\delta\nu|^{3/2}$ when doping the transition point, which implies that $\kappa \sim 1/ \left(\frac{\partial^2 E}{\partial \delta \nu ^2}\right) \sim |\delta \nu|^{1/2}$ has the same scaling behavior as $\kappa_{x,y}$. This agrees with the expectation that the system now has emergent conformal invariance, and suggests that the sound velocities of the superfluid $v_{\text{s},i} = \sqrt{\kappa_i/\kappa}$ remain finite even in the dilute limit when doping the critical point.

\section{The role of disorder}
\label{app: disorder}

As in all QH systems, the disorder in our problem will stablize the QH state against a finite density of pseudospin and form a QH plateau. In this section we estimate the width of the plateau--that is, the critical pseudospin density for the transition into the ES state--near the topological transition.

\subsection{The critical doping concentration}

In the presence of weak, local disorder in the pseudospin channel, we expect the effective LLs of $F$ to be broadened in energy and most of the states therein will be localized and the Hall response in each LL will be carried by only one extended state~\cite{Laughlin_extended,khmelnitskii1983quantization,PhysRevB.32.1311}. The strength of disorder is characterized by a dimensionless parameter $\Omega_\text{cl}\tau \propto |B_\text{eff}|$ which is proportional to the effective magnetic field strength $B_\text{eff}$ felt by $F$. In the current setup, both the density of $F$ and $B_\text{eff}$ are controlled by the pseudospin density $\rho_\text{s}$. Therefore, doping the pseudospin is equivalent to scaling the disorder strength. More concretely, increasing the pseudospin density effectively drives the $F$ system into the clean limit, during which the extended states will descent and pass the Fermi level one by one (the Chern number $C_F$ changes accordingly), eventually leading to the ES state we constructed in the main text.

For the current case, the relevant dispersion of $F$ is Dirac, and for a single Dirac fermion with small Dirac mass, Ref.~\cite{PhysRevB.90.165435} has obtained the following equation for the position (chemical potential) of the $n_0(\geq 0)$-th LL in the presence of disorder:
\begin{align}
    \mu  = \sqrt{n_0 2 B v_\text{F}^2 \frac{1+(\Omega_\text{cl}\tau)^2}{(\Omega_\text{cl}\tau)^2}+m_\text{D}^2}
\end{align}
where 
\begin{align}
    \Omega_\text{cl}\tau = \frac{B v_\text{F}^4}{n_\text{imp} V_\text{imp}^2}\frac{4}{\mu^2 +3 m_\text{D}^2}.
\end{align}
Approximately relating the density $\rho$ and $\mu$ with the $B=0$ result, 
\begin{align}
    \mu^2 = m_\text{D}^2 + 4\pi v_\text{F}^2 \rho 
\end{align}
we find the critical density at which the extended state resides at the Fermi level is determined by
\begin{align}
    \rho = \frac{B}{2\pi} n_0\frac{1+(\Omega_\text{cl}\tau)^2}{(\Omega_\text{cl}\tau)^2}
\end{align}
which gives good approximation to the exact result in both the strong and weak disorder limit. We therefore proceed with this simple expression for the sake of clarity. 

For the FP system $F$, the effective $B_\text{eff}\propto\rho_\text{s}$ (assuming $\rho_\text{s}>0$ without loss of generality). Substituting this relation into the above equation, we find the simple criterion for the transition density into the ES state:
\begin{align}
    \Omega_\text{cl}\tau  = c
\end{align}
where $c$ is an $\mathcal{O}(1)$ constant that depends on the detailed nature of the transition. The critical density thus admits an estimate:
\begin{align}
    \rho_{\text{s,critical}} = \frac{m_D^2}{c'v_\text{F}^4/(n_\text{imp} V_\text{imp}^2) -\pi v_\text{F}^2}
\end{align}
where $c' $ is another $\mathcal{O}(1)$ constant. Thus we expect that the ES state to emerge immediately upon doping the critical points as long as the disorder is weak enough. 

\subsection{The spin conductivity upon transitioning into the ES phase}

This plateau transition picture implies that the spin conductivity tensor admits universal behavior at the transition point into the ES state. To see that, we first discuss the physical response of the electrons at mean-field level in terms of the resopnse of the anyons fields, $F$, as follows. 

We will assume weak fluctuations of the emergent gauge fields, such that the response of the $F$ fields to the gauge fields can be effectively captured by the corresponding polarization tensor $\Pi^{\mu\nu}_F = \langle J^\mu J^\nu\rangle_F$ in a way that the effective action is quadratic in the fluctuating part of the gauge fields $\delta a$. Then the effective response theory within the spin sector respectively reads (derived from Eq.~\ref{app: eq: L CS'}):
\begin{align}
    \mathcal{L}& =  \frac{1}{2} \delta a_\mu \Pi^{\mu\nu}_F \delta a_\nu +\frac{2n}{4\pi} \delta a \mathrm{d}\delta a  -\frac{2n}{2\pi} \delta a \mathrm{d} A^\text{s} +\frac{2}{4\pi}A^\text{s}\mathrm{d} A^\text{s}
\end{align}
For simplicity, we then take the temporal gauge for all the gauge fields ($a_0=A_0=0$). Then further integrating out $\delta a$ yields the following expression for the electrons' spin conductivity in terms of that of the $F$ conducitivity:
\begin{align}
   2\pi \hat{\sigma}^\text{s}(\bm{q},\omega) = 
   2\pi  \Pi^\text{s}(\bm{q},\omega) / (\mathrm{i}\omega) &= 2n E - 4n^2\mathrm{i}\omega E \cdot [2\pi \Pi_F + 2n \mathrm{i}\omega E]^{-1}\cdot E \nonumber\\
   &= 2n E - 4n^2 E \cdot [2\pi\hat{\sigma}_{F} + 2n E]^{-1}\cdot E 
\end{align}
where
\begin{align}
    E \equiv \begin{bmatrix}
        0 & 1 \\
        -1 & 0
    \end{bmatrix}
\end{align}
In terms of resistivity, this relation can be simplified:
\begin{align}
    \frac{1}{2\pi}\hat{R} =  \frac{1}{2\pi} \hat{R}^F + \frac{1}{2n} \begin{bmatrix}
        0 & -1 \\
        1 & 0
    \end{bmatrix}
\end{align}

We assume that the transition into the ES state is driven by one extended state passing the Fermi level of $F$, i.e. happens when the Chern number of $F$, $C_F$, changes from $-(2n-1)$ to $-2n$. In this scenario, we expect
\begin{align}
    2\pi \hat{\sigma}_{F,ij} = & \frac{1}{2} \delta_{ij} - \frac{4n-1}{2} \epsilon_{ij  }
\end{align}
where $I$ is the identity. This means that the electron conductivity at this critical point to the ES state is
\begin{align}
    2\pi \hat{\sigma}^\text{s}_{ij} = 4n^2 \delta_{ij} - (4n^2-2n) \epsilon_{ij}.
\end{align}

\end{document}